\documentclass[10pt,conference]{IEEEtran}
\IEEEoverridecommandlockouts
\usepackage{graphicx}
\usepackage{booktabs}
\usepackage{amsmath}
\usepackage{rotating}
\usepackage{algorithm}
\usepackage{algpseudocode}
\usepackage{xspace}
\usepackage{multirow}
\usepackage{url}
\usepackage{amssymb}
\usepackage{mathtools}
\usepackage{balance} 
\usepackage{tikz}
\usetikzlibrary{calc}
\usepackage{makecell}
\usepackage[most]{tcolorbox}
\usepackage{pgfplots}
\pgfplotsset{compat=1.18}
\usepackage{listings}
\usepackage{xcolor}
\usepackage{url}
\usetikzlibrary{tikzmark}
\usepackage{lstlinebgrd}
\usepackage[hidelinks]{hyperref}

\newcommand{\tool}{\textsc{VulnSEER}\xspace}

\tcbset{
  colback=gray!15,    % 设置背景为更浅的灰色（gray!10 表示 10% 灰色）
  colframe=gray!20,   % 设置边框颜色为稍深的灰色
  arc=1mm,            % 设置圆角半径为 5mm
  boxrule=0.5pt,      % 设置边框线的粗细
  width=\linewidth,   % 设置框的宽度为页面的宽度
  left=5pt,           % 设置框与文字的左边距
  right=5pt,          % 设置框与文字的右边距
  top=5pt,            % 设置框与文字的上边距
  bottom=5pt          % 设置框与文字的下边距
}

\definecolor{codeblue}{RGB}{0,38,150}
\definecolor{codeteal}{RGB}{0,120,150}
\definecolor{codestring}{RGB}{0,120,40}
\definecolor{codeanno}{RGB}{135,145,30}
\definecolor{codecomment}{RGB}{130,130,130}
\definecolor{pathgreen}{RGB}{176,215,145}
\definecolor{sinkred}{RGB}{236,137,150}

\lstdefinestyle{motivatingjava}{
  language=Java,
  basicstyle=\ttfamily\scriptsize,
  keywordstyle=\color{codeblue}\bfseries,
  stringstyle=\color{codestring},
  commentstyle=\color{codecomment},
  morekeywords={Nonnull,Nullable,OutputStream,IOException,SQLException,String,Table,Config,Filer,Streamer,Coder},
  keywordstyle=[2]\color{codeteal},
  morekeywords=[2]{TableUtils,DecoratedFiler,StreamerFiler,ChainedStreamer,CompressStreamer,SnappyCoder,SnappyOutputStream},
  numbers=left,
  numberstyle=\scriptsize\color{black},
  numbersep=8pt,
  xleftmargin=1.8em,
  frame=none,
  breaklines=true,
  columns=fullflexible,
  keepspaces=true,
  showstringspaces=false,
  aboveskip=0pt,
  belowskip=0pt,
  lineskip=-1.5pt,
  escapeinside={(*@}{@*)},
  linebackgroundcolor={%
    \ifnum\value{lstnumber}=11\color{pathgreen}\fi
    \ifnum\value{lstnumber}=19\color{pathgreen}\fi
    \ifnum\value{lstnumber}=29\color{pathgreen}\fi
    \ifnum\value{lstnumber}=41\color{pathgreen}\fi
    \ifnum\value{lstnumber}=53\color{pathgreen}\fi
    \ifnum\value{lstnumber}=60\color{sinkred}\fi
    \ifnum\value{lstnumber}=61\color{sinkred}\fi
  }
}

\definecolor{vulnred}{RGB}{170,0,0}

\lstdefinestyle{layerctx}{
  basicstyle=\ttfamily\footnotesize\color{vulnred},
  numbers=left,
  numberstyle=\tiny\color{black},
  numbersep=8pt,
  xleftmargin=1.8em,
  frame=none,
  breaklines=true,
  columns=fullflexible,
  keepspaces=true,
  showstringspaces=false,
  aboveskip=0pt,
  belowskip=0pt,
  lineskip=-1.5pt
}
\def\BibTeX{{\rm B\kern-.05em{\sc i\kern-.025em b}\kern-.08em
    T\kern-.1667em\lower.7ex\hbox{E}\kern-.125emX}}
\begin{document}

\title{\tool: Layer-Wise Exploit State Re-expression for Vulnerable-Dependency Exploit Payload Synthesis}

\author{
\IEEEauthorblockN{\makebox[0.33\textwidth]{Likai Ye}}
\IEEEauthorblockA{
\textit{School of Software Technology}\\
\textit{Zhejiang University}, Hangzhou, China\\
\textit{The University of Melbourne}\\
Melbourne, Australia\enspace yelikai@zju.edu.cn
}
\and
\IEEEauthorblockN{\makebox[0.33\textwidth]{Jiamou Sun}}
\IEEEauthorblockA{
\textit{Technology, CSIRO}\\
Canberra, Australia\\
frank.sun@csiro.au
}
\and
\IEEEauthorblockN{\makebox[0.33\textwidth]{Tingting Bi}}
\IEEEauthorblockA{
\textit{The University of Melbourne}\\
Melbourne, Australia\\
Tingting.Bi@unimelb.edu.au
}
\and[\hfill\mbox{}\par\noindent\mbox{}\hfill]
\IEEEauthorblockN{\makebox[0.33\textwidth]{Zhenchang Xing}}
\IEEEauthorblockA{
\textit{Technology, CSIRO}\\
Canberra, Australia\\
zhenchang.xing@csiro.au
}
\and
\IEEEauthorblockN{\makebox[0.33\textwidth]{Lingfeng Bao}}
\IEEEauthorblockA{
\textit{Zhejiang University}\\
\textit{Hangzhou High-Tech Zone (Binjiang)}\\
\textit{Institute of Blockchain and Data Security}\\
Hangzhou, China\\
lingfengbao@zju.edu.cn
}
\and
\IEEEauthorblockN{\makebox[0.33\textwidth]{Xiaoxue Ren}}
\IEEEauthorblockA{
\textit{School of Software Technology}\\
\textit{Zhejiang University}\\
\textit{Hangzhou High-Tech Zone (Binjiang)}\\
\textit{Institute of Blockchain and Data Security}\\
Hangzhou, China\enspace xxren@zju.edu.cn
}
}

% \and
% \and
% \and
% \and
% \and

\maketitle
\begingroup\renewcommand\thefootnote{}\footnotetext{* Corresponding author: Xiaoxue Ren (xxren@zju.edu.cn).}\addtocounter{footnote}{-1}\endgroup

\begin{abstract}
Modern Java applications widely reuse third-party libraries, but the presence or reachability of a vulnerable dependency does not necessarily imply that the client project is exploitable. Determining exploitability requires synthesizing an entry-side payload that drives a project-side entry function to a vulnerable library sink and satisfies the vulnerability-specific oracle. Existing test migration, state mimicry, and directed fuzzing techniques become less effective on long client-to-library call chains because the vulnerability-triggering condition is often represented by different caller-controllable states across layers and governed by off-chain construction or propagation logic. 
We present \tool, an LLM-guided exploit state re-expression framework for vulnerable-dependency exploit payload synthesis. \tool derives a sink-level requirement from a library-side exploit sketch, infers layer-wise exploit goals backward from the vulnerable sink to the project-side entry, reorganizes them into a forward execution plan, and uses RL-based exploit-relevant context selection to guide exploit state resolution. The resolved argument and object-state bindings are assembled into an executable payload and validated through instrumented replay. On 269 ground-truth exploitable call chains from 49 Java projects and 32 vulnerabilities, \tool achieves a mean overall success rate of 87.11\% ($\pm$0.62\%), outperforming MAGNETO, VESTA, TRANSFER, and VULEUT, which achieve 58.74\%, 16.36\%, 19.70\%, and 17.84\%, respectively.

\end{abstract}


\section{Introduction}
\label{sec:intro}
Modern Java applications heavily rely on third-party libraries to accelerate development, but these dependencies also introduce security and maintenance risks~\cite{sun2023silent}. When a vulnerable dependency is reported, developers must determine not only whether the dependency is exploitable in their application, but also whether the security update is compatible with the existing dependency environment and how much engineering effort is required for mitigation. However, a vulnerable dependency does not necessarily make the client application exploitable~\cite{ponta2020detection,mir2023effect}. Exploitability requires an executable path from a project entry to the vulnerable upstream sink that satisfies the vulnerability-specific triggering condition. Therefore, the key challenge is exploitability assessment rather than vulnerable dependency detection alone.

%Modern Java applications heavily rely on third-party libraries to accelerate development and reuse mature functionality~\cite{sun2023silent}. However, these libraries also introduce security and maintenance risks. When a vulnerable dependency is reported, developers need to know not only whether the dependency is vulnerable, but also whether a security update is compatible with the existing dependency environment and how much engineering effort is required to mitigate a potentially unfamiliar threat.In fact, depending on a vulnerable library does not necessarily imply that the client application is exploitable~\cite{ponta2020detection,mir2023effect}.A dependency vulnerability becomes practically relevant only when the client project contains an executable path that starts from a project-side entry function, propagates through client-side function call chains, reaches the vulnerable sink of upstream library (i.e., the vulnerable upstream function used by the client project), and satisfies the vulnerability-specific triggering condition. Therefore, the central problem is not vulnerable dependency detection alone, but exploitability assessment, namely determining whether an entry-side payload can drive the client project to trigger the vulnerable library behavior.

Existing studies on vulnerable dependencies mainly focus on vulnerability existence and reachability analysis. The former identifies vulnerable dependencies or code patterns~\cite{pashchenko2022vuln4real,zhao2023software,imtiaz2021comparative,li2022sysevr,fu2022linevul}, while the latter determines whether vulnerable library functions are reachable from client code~\cite{ponta2018beyond,zhang2024does,jia2025impact,wang2025savant,ma2024vulnet}. However, neither establishes whether a client-side entry can construct and propagate the runtime state required to trigger the vulnerability. Recent work therefore focuses on exploitability validation by generating executable tests or exploits for downstream projects, 
% including test migration approaches (e.g., \textsc{Transfer}~\cite{kang2022test} and \textsc{Vesta}~\cite{chen2024exploiting}) and Large Language Model (LLM)-assisted directed fuzzing (e.g., \textsc{Magneto}~\cite{zhou2024magneto}).
%RA-Q1 & Comment 2
including test migration approaches (e.g. \textsc{Transfer}~\cite{kang2022test} and
\textsc{Vesta}~\cite{chen2024exploiting}), test generation guided by reachable path analysis (e.g. \textsc{VulEUT}~\cite{gao2025vulnerability}), and large language model (LLM)-assisted directed fuzzing (e.g.\textsc{Magneto}~\cite{zhou2024magneto}).

%Existing studies on vulnerable dependencies mainly focus on vulnerability existence analysis and vulnerability reachability analysis. 
%The former identifies whether a project depends on vulnerable libraries or contains vulnerable code patterns~\cite{pashchenko2022vuln4real,zhao2023sca,imtiaz2021comparative,li2021sysevr,fu2022linevul}, while the latter determines whether vulnerable functions can be reached from client code~\cite{ponta2018beyond,zhang2024vascanner, jia2025impact,wang2025savant, ma2024vulnet}.
%However, neither line of work establishes whether a client-side entry can construct and propagate the runtime state required to trigger the vulnerable library behavior. 
%Recent work has therefore shifted toward vulnerability exploitation analysis, which aims to generate executable tests or exploits for validating vulnerable-dependency exploitability in downstream projects.
%LLM-based test migration and test mimicry techniques, such as \textsc{Transfer}~\cite{kang2022test} and \textsc{Vesta}~\cite{chen2024vesta}, adapt library-side vulnerability tests or public exploits to client projects, and LLM-assisted directed fuzzing techniques, such as \textsc{Magneto}~\cite{zhou2024magneto}. 

However, our preliminary study (Section~\ref{sec:motivate}) shows that existing techniques degrade substantially as client-side call chains become longer. For chains of length at least six, the SoTA approach \textsc{Magneto} succeeds on only 1 of 16 cases (6.25\%), while \textsc{Vesta} and \textsc{Transfer} solve none. We attribute these failures to two challenges. First, the vulnerability-triggering \emph{exploit state} may reside in hidden program states scattered across function call layers rather than explicit call arguments, such as receiver fields, factory decisions, or configuration entries. Exploit synthesis must therefore identify which state at the current layer governs the downstream condition and translate the condition into constraints over that state (\textbf{Challenge-1}).  Second, exploit-state construction and propagation often occur outside the main call chain, where indiscriminate context expansion introduces substantial noise. Consequently, it is essential to select the program context that governs exploit state propagation (\textbf{Challenge-2}). Although related problems have been studied in test generation, symbolic execution, and program slicing~\cite{baldoni2018survey,tu2026cottontail,sridharan2007thin}, existing LLM-based exploit synthesis techniques do not jointly model these challenges, limiting their effectiveness on long client-to-library call chains.

Based on this insight, we propose \tool, an LLM-guided exploit state re-expression with exploit-relevant context selection framework for client-level exploit payload synthesis in Java projects with vulnerable third-party dependencies.
Given a downstream project and a target vulnerability specification, \tool extracts vulnerability-related information and exploit requirements from both the client project and the vulnerable library. 
To address Challenge-1, \tool performs backward exploit-goal inference, re-expressing sink-level triggering conditions over caller-visible program states.  
The inferred goals are then reorganized into an entry-to-sink execution plan, which specifies what each layer must resolve during forward exploit state construction. 
To address Challenge-2, \tool uses a reinforcement learning (RL)-based context selector to retain only exploit-relevant context evidence for LLM-guided exploit state resolution. 
Finally, \tool assembles the resolved argument bindings and object-state bindings into an executable payload and validates it through execution against the vulnerability-specific oracle.

We evaluate \tool on an expanded benchmark containing 269 
ground-truth exploitable call chains from 49 Java client projects and 32 upstream vulnerabilities. 
% Under GPT-5.2, \tool achieves an overall success rate of 87.73\%, outperforming \textsc{Magneto} (58.74\%), \textsc{Vesta} (16.36\%), and \textsc{Transfer} (19.70\%). The advantage is more pronounced on long chains, where \tool solves 12 of 18 cases, compared with 1 for \textsc{Magneto} and none for \textsc{Vesta} or \textsc{Transfer}.
%RA-Q1 & Comment 2
Under GPT-5.2, \tool achieves a mean overall success rate of 87.11\% ($\pm$0.62\%), outperforming \textsc{Magneto} (58.74\%), \textsc{Vesta} (16.36\%),
\textsc{Transfer} (19.70\%), and \textsc{VulEUT} (17.84\%). The advantage is more pronounced on long chains, where \tool solves 12 of 18 cases, compared with 1 for \textsc{Magneto} and none for \textsc{Vesta}, \textsc{Transfer}, and \textsc{VulEUT}.
Ablation studies further show that both backward exploit-goal inference and exploit-relevant context selection are critical, reducing the success rate to 56.88\% and substantially degrading long-chain performance when removed or replaced.

This paper makes the following contributions:

\begin{itemize}
    \item We conduct a preliminary study of long-chain exploit synthesis and identify two recurring challenges. These include re-expressing sink-level exploit conditions over caller-controllable state and selecting exploit-relevant project context.

    \item We propose \tool, an LLM-guided exploit synthesis framework combining backward exploit-goal inference with an RL-based context selector. It further validates synthesized exploits through vulnerability-specific oracles.

    \item We conduct extensive experiments on our extended benchmark, covering diverse exploitable call chains across multiple Java projects and vulnerabilities, and demonstrate that \tool achieves superior performance over existing approaches.

    % \item The artifacts of \tool are available at \url{https://github.com/Ylkylkylk/VulnSEER}.
\end{itemize}

% \section{Motivation}
% \input{section/02motivating_example}
% \label{sec:motivate}

\section{Preliminary Study}
\label{sec:motivate}

To understand where existing techniques fail, we study how exploit-synthesis effectiveness varies with call-chain length, asking whether they can still generate a verified entry-side payload as the vulnerable sink becomes farther from the project entry by multiple caller-callee boundaries.

%To quantify where existing techniques fail, we conduct a preliminary study on how their effectiveness changes with respect to call-chain length. The key question is whether a method can still synthesize a verified entry-side payload when the vulnerable sink is separated from the project entry by multiple caller-callee boundaries.
\subsection{Experimental Setup}
%\noindent\textbf{Experimental Setup.}

\subsubsection{Problem Formulation}
Given a Java project $\mathcal{P}$, a vulnerable sink $s$ in a third-party dependency, a vulnerability-specific oracle $\mathcal{O}$, and an entry-to-sink call chain
$\rho = \langle N_0 \xrightarrow{c_0} \cdots \xrightarrow{c_{L-1}} N_L \rangle$
with $N_0$ the project entry and $N_L=s$, each method aims to generate an entry-side payload that satisfies $\mathcal{O}$.

%Let $\mathcal{P}$ denote a Java project with resolved third-party dependencies, $s$ a vulnerable sink in a dependency, and $\mathcal{O}$ a vulnerability-specific oracle. Given an entry-to-sink call chain, each method aims to generate an entry-side payload that reaches $s$ and satisfies $\mathcal{O}$. We represent such a vulnerability-reaching path as
%$\rho = \langle N_0 \xrightarrow{c_0} N_1 \xrightarrow{c_1} \cdots \xrightarrow{c_{L-1}} N_L \rangle$, where $N_0$ is a project-side entry node, $N_L = s$, and each $c_k$ is a call site in $N_k$ that may invoke $N_{k+1}$.

\subsubsection{Baselines}
We compare three representative existing methods that cover the main technical directions for vulnerable-dependency exploit generation:
\begin{itemize}
    \item \textbf{\textsc{Magneto}}~\cite{zhou2024magneto} represents fuzzing-based exploit generation. It decomposes a call chain into caller--callee steps, uses an LLM to generate input seeds for each step in reverse direction, and applies directed fuzzing to preserve the exploit-relevant state along the chain.
    \item \textbf{\textsc{Vesta}}~\cite{chen2024exploiting} represents migration-based test generation. It generates or reuses a client-side test reaching the vulnerable function and migrates concrete triggering values from a public exploit into that test using predefined transformation rules.
    \item \textbf{\textsc{Transfer}}~\cite{kang2022test} represents state-mimicry-based test generation. It extracts the triggering program state from a library-side vulnerability-triggering test and guides evolutionary search to reproduce a similar state from the client project entry.
\end{itemize}
Together, these baselines cover three principal strategies for transferring library-side vulnerability knowledge to client-side exploits. 
%RB_Q2&comment7
Although \textsc{VulEUT}~\cite{gao2025vulnerability} is a recently proposed client-side exploit generation approach, it relies on rule-based heuristics to handle parameter propagation and type conversions. Such limited modeling may impair its effectiveness in handling complex exploit chains and lead to a high false-positive rate. Therefore, we exclude \textsc{VulEUT} from the preliminary study.

Recently, there are also some newly proposed works including \textsc{VulnSage}~\cite{chen2026multiagent} and \textsc{Explode.js}~\cite{marques2025automated}. However, their task formulations differ from ours: \textsc{VulnSage} targets alert-driven exploit generation without considering third-party dependencies, while \textsc{Explode.js} synthesizes multi-interaction exploits but only supports Node.js libraries. 
%RA-Q2 & Comment 3：ChainFuzz
\textsc{ChainFuzz}~\cite{deng2025chainfuzz} similarly adapts upstream PoCs through taint-guided byte-level mutation, but its LLVM/AFLGo infrastructure targets C/C++ rather than Java.
We therefore discuss these systems as related work, but do not include them as experimental baselines.

\subsubsection{Benchmark}
We use the \textsc{Magneto} benchmark, which contains 32 real-world Java vulnerabilities, 21 vulnerable dependency libraries, 49 client projects, and 182 ground-truth exploitable call chains. Each case provides a vulnerable dependency, an entry-to-sink call chain, and a manually curated oracle test for validating exploitability.

\subsubsection{Metrics}
We use the chain-level success rate: $\mathrm{SR} = \frac{N_{\mathrm{success}}}{N_{\mathrm{gt}}}$ as the primary metric,
where $N_{\mathrm{gt}}$ denotes the number of ground-truth exploitable call chains and $N_{\mathrm{success}}$ is the number for which a verified exploit is generated. A chain is counted as successful only if the generated exploit triggers the corresponding oracle.

\subsubsection{Implementation}
We run our experiments on a Linux server equipped with two Intel(R) Xeon(R) Platinum 8358P CPUs @ 2.60GHz, 2.0 TiB of RAM, and eight NVIDIA A800 80GB PCIe GPUs, running Ubuntu 20.04.4 LTS with Linux kernel 5.4.0. All LLM calls use temperature 0.2. We use the backend default maximum output length, retry failed API calls or malformed JSON responses up to four times with a 2-second linear backoff.

\begin{figure}[htbp]
    \centering
    \includegraphics[width=0.48\textwidth]{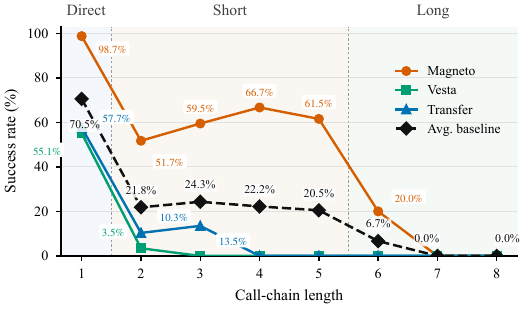} \vspace{-3mm}
    \caption{Baseline success-rate trends across call-chain lengths.} \vspace{-3mm}
    \label{fig:baseline-trend}
\end{figure}

\subsection{Results}
\subsubsection{Barriers of long-chain}
Fig.~\ref{fig:baseline-trend} summarizes the success rates of the three baselines across different call-chain lengths. 
Two clear performance transitions can be observed.
First, the success rates of all baselines decrease sharply from length-1 chains to length 2 (98.7\%, 55.1\%, and 57.7\% vs.\ 51.7\%, 3.5\%, and 10.3\% for \textsc{Magneto}, \textsc{Vesta}, and \textsc{Transfer}, respectively), which shows baselines are more effective when the client directly invokes the vulnerable function than when exploit generation requires even one additional caller--callee propagation step.
Second, another sharp decline occurs at length 6, where the average success rate falls from 22.2\% (lengths 2--5) to 6.7\%, with no baseline succeeding on lengths 7 or 8.
Overall, among the 16 cases with length at least 6, \textsc{Magneto} solves only one, while \textsc{Vesta} and \textsc{Transfer} solve none. Based on these two empirical transition points, we refer to length-1 paths as \emph{direct calls}, lengths 2--5 as \emph{short chains}, and lengths at least 6 as \emph{long chains}.
%The average success rate drops sharply when moving from length~1 to length~2. This indicates that the first major difficulty arises once exploit generation moves beyond direct sink invocation and requires interprocedural propagation through client code. More importantly, among non-direct cases, the decline becomes most pronounced when the chain length reaches~6, suggesting that existing techniques become increasingly ineffective when the exploit condition must be preserved across several caller-callee boundaries. Following this trend, we treat length~1 as direct calls, lengths~2--5 as short chains, and lengths at least~6 as long chains.

\begin{figure*}[htbp]
    \centering
    \includegraphics[width=1\textwidth]{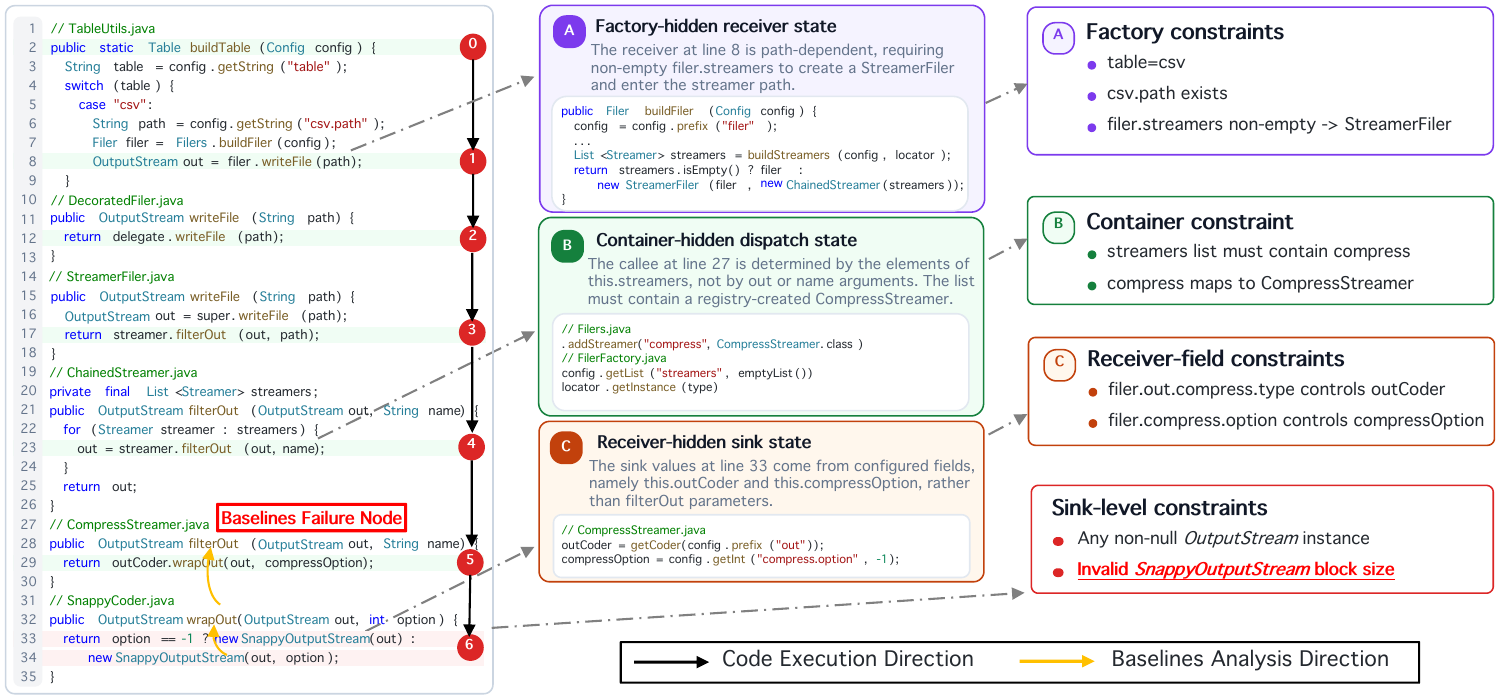} \vspace{-3mm}
    \caption{Exploit state propagation in a representative long-chain case.}
    \label{fig:flow-motivating}
    \vspace{-3mm}
\end{figure*}

\subsubsection{Failure mechanism in long-chain cases}
These results identify where existing methods deteriorate, but chain length alone does not explain why. We therefore inspect their long-chain failures and use the representative \href{https://github.com/lithiumtech/flow/commit/ac480b51e1ca1d3bcc2dcafe202c36493a0db0ac}{\textcolor{blue}{\texttt{flow}}} case, which depends on a vulnerable version of \texttt{snappy-java} (\href{https://nvd.nist.gov/vuln/detail/CVE-2023-43642}{\textcolor{blue}{CVE-2023-43642}}) and cannot be exploited by any baseline, to illustrate the underlying difficulty.
%The \texttt{flow} project depends on a vulnerable version of \texttt{snappy-java} affected by \href{https://nvd.nist.gov/vuln/detail/CVE-2023-43642}{\textcolor{blue}{CVE-2023-43642}}. None of the three baselines generates a validated entry-side exploit for this case. 
%At the vulnerable sink, exploitation requires invoking the two-argument
%\texttt{SnappyOutputStream} constructor with a non-null
%\texttt{OutputStream} and a vulnerability-triggering block-size value. However, these sink arguments are not directly propagated from the project entry.
%\likai{Motivated by the above observation, we further inspect a representative long-chain case to understand why existing techniques fail. The example is drawn from the \texttt{flow}\footnote{\url{https://github.com/lithiumtech/flow}} project, which depends on a vulnerable version of \texttt{snappy-java} affected by \href{https://nvd.nist.gov/vuln/detail/CVE-2023-43642}{\textcolor{blue}{CVE-2023-43642}}. In this case, none of the three baselines generates a validated entry-side exploit. As shown in Fig.~\ref{fig:flow-motivating}, the entry reads configuration values, constructs a \texttt{Filer}, and invokes \texttt{filer.writeFile(path)}. The stream then passes through filer decorators and streamer abstractions before \texttt{CompressStreamer} delegates to a configured coder, and \texttt{SnappyCoder} finally selects the vulnerable constructor.}
As shown in Fig.~\ref{fig:flow-motivating}, although the vulnerable sink (\texttt{SnappyOutputStream} constructor) requires only a valid \texttt{OutputStream} and a vulnerability-triggering block size, these values are not directly propagated from the project entry. 
The block-size is stored in the \texttt{compressOption} field of the \texttt{CompressStreamer} receiver, while the \texttt{SnappyCoder} implementation is determined by the receiver field \texttt{outCoder}. Reaching this receiver further requires the \texttt{streamers} container to be non-empty and contain a \texttt{CompressStreamer}; otherwise, execution skips compression or dispatches to another streamer. Moreover, neither these receiver fields nor the container contents are explicit in the main call chain, but are constructed from \texttt{Config} through the filer factory, streamer registry, and \texttt{CompressStreamer} constructor in off-chain logic. This exposes two longstanding program-analysis challenges that remain largely unmodeled by existing LLM-based exploit synthesis approaches.

\noindent\textbf{Challenge-1: Re-expressing exploit conditions over layer-controllable state.}
At each caller--callee boundary, explicit call arguments may not determine whether the downstream exploit condition is satisfied. Instead, the relevant state may reside in receiver fields, container elements, factory decisions, or configuration entries. Exploit synthesis must therefore identify the caller-controllable state governing the downstream behavior and re-express the exploit condition over that state.

%At each caller--callee boundary, the variables explicitly passed at the call site are not necessarily the state that determines whether the downstream exploit condition is satisfied. A sink argument may be derived from a receiver field, a container element, a factory decision, or a configuration entry rather than from a caller argument with the same value. Exploit synthesis must therefore identify which state visible and controllable at the current layer governs the required downstream behavior, and re-express the exploit condition over that state.

In the \texttt{flow} case, the block-size requirement at the vulnerable constructor must first be re-expressed as a constraint on
\texttt{CompressStreamer.compressOption}, while the required coder becomes a constraint on
\texttt{CompressStreamer.outCoder}. At the preceding layer, these receiver-field requirements become constraints on the contents and runtime types of elements in the
\texttt{FilerFactory.streamers} container. At the project entry, they are finally expressed as configuration entries that cause the factory and registry to construct the required wrapper graph.

Existing techniques do not explicitly derive this layer-specific correspondence. \textsc{Vesta} migrates concrete exploit values, which is effective when the triggering values remain direct or simply transformed arguments. \textsc{Transfer} searches for a program state resembling the library-side vulnerability witness, but does not explicitly determine which caller-controllable fields, containers, or factory choices represent that state at preceding layers. \textsc{Magneto} propagates exploit information between adjacent call steps, but its seed generation and fuzzing do not explicitly maintain a structured correspondence between the downstream condition and the different state representation visible at each caller. The challenge is therefore not merely to carry a dangerous value through a call chain. It is to determine, at every layer, which caller-controllable state represents the downstream exploit condition.
%\likai{Existing baselines mainly rely on local transfer of sink-level triggering information. 
%For example, \textsc{Magneto} performs step-wise directed fuzzing and propagates exploit information between adjacent call steps, while \textsc{Vesta} and \textsc{Transfer} reuse vulnerability-triggering values extracted from library-side tests. 
%However, such local transfer becomes fragile once the sink-level condition is re-encoded into project-specific state. In Fig.~\ref{fig:flow-motivating}, the sink requires the two-argument \texttt{SnappyOutputStream} constructor to receive a vulnerability-triggering block-size value. When reasoning one layer upward to \texttt{CompressStreamer.filterOut}, this value is no longer an explicit argument of the current method. Instead, it is stored in the receiver field \texttt{compressOption}, which is initialized elsewhere from configuration logic. 
%Therefore, baselines still recover or follow the call from \texttt{filterOut} to \texttt{wrapOut}, but fail to construct the receiver state required to pass the correct \texttt{option} value to the sink. This problem compounds across longer chains. Thus, the challenge is not merely to pass a dangerous value through a call chain, but to preserve its exploit semantics as the representation changes from sink arguments to receiver fields, containers, factory decisions, and configuration entries. }

\noindent\textbf{Challenge-2: Selecting context that governs exploit-state propagation.}
Even after identifying the exploit-relevant state at a layer, the call chain alone is often insufficient to determine whether that state reaches the next layer with the required value and type. The necessary evidence may reside outside the direct call chain, such as in factories, constructors, helper methods, or dynamic-dispatch logic. The key challenge is therefore to identify the context governing how exploit-relevant state is constructed, propagated, or selected. For example, in the \texttt{flow} case, the main call chain does not explain why \texttt{FilerFactory.streamers} contains a \texttt{CompressStreamer} or why its fields resolve to a vulnerability-triggering \texttt{SnappyCoder} and option, as these are determined by off-chain factory, registry, configuration, and constructor logic.

Another example appears in \href{https://github.com/webbit/webbit/commit/5403edae6373ea4ff47cfccda0d61135e77db5b4}{\textcolor{blue}{\texttt{webbit}}}, where the wrapper method \texttt{cookieValue(name)} eventually invokes Netty's vulnerable \texttt{decode(headerValue)} through request delegation and inbound-cookie parsing. Unlike the \texttt{flow} case, the exploit-relevant state is not substantially transformed. It remains the attacker-controlled \texttt{Cookie} header string. The key difficulty is not determining a new state carrier at each layer, but recovering the context that proves how this value is propagated. The wrapper delegates to the request, which retrieves the inbound header via \texttt{headers(COOKIE\_HEADER)} and forwards it to \texttt{decode}. Without selecting these fragments, an analysis may incorrectly treat the explicit \texttt{name} argument of \texttt{cookieValue} as the sink input.
%identifying the context that exposes this state carrier, because the wrapper delegates to the underlying request, the Netty request implementation reads the header through \texttt{headers(COOKIE\_HEADER)}, and the parser forwards each header value to the decoder. Without this context, an analysis may follow the call chain but miss that the decoder input is governed by the inbound header rather than the \texttt{name} argument of \texttt{cookieValue}.
%\likai{Preserving vulnerability-triggering semantics requires more than following the highlighted call-chain lines. Once the condition changes representation, the analysis must inspect the off-chain code that explains how the required state is constructed. 
%定义context
%For example, line~8 alone cannot explain why the receiver may become a \texttt{StreamerFiler}; the relevant context is the filer factory that wraps the base \texttt{Filer} only when the configured streamer list is non-empty. Similarly, line~33 alone cannot explain why \texttt{outCoder} resolves to \texttt{SnappyCoder} or why \texttt{compressOption} has a vulnerability-triggering value; the relevant context is the constructor and registry logic that maps configuration keys to these receiver fields.}

The three baselines expose complementary limitations in recovering such context. \textsc{Magneto} focuses on local call-site information but may miss off-chain construction or transformation logic, while \textsc{Transfer} and \textsc{Vesta} recover exploit-relevant values without systematically identifying the client-side code that constructs, modifies, or preserves them.
%RB_Q1&comment1: Statistics of Baseline Failures
To further investigate these limitations, we conduct a fine-grained failure analysis of \textsc{Magneto}, the strongest baseline, whose failures can be attributed to concrete causes. Among 47 analyzed failures, 38 involve missing cross-layer exploit-state re-expression, 34 lack off-chain propagation context, five involve uncontrollable third-party API responses, and one is blocked by a runtime exception. These categories are not mutually exclusive, and the first two recur across most cases.
However, simply including more surrounding code is ineffective, as projects often contain many syntactically nearby but irrelevant factories, implementations, accessors, transformations, and helper methods. In our benchmark, long call chains contain 36 candidate context items on average and up to 179. The challenge is therefore to identify the code fragments that semantically govern the exploit-relevant state while excluding unrelated nearby context.

%The three baselines expose complementary limitations in recovering such context. \textsc{Magneto} extracts focal information around individual call steps, but may omit off-chain construction or transformation logic and may include alternative implementations unrelated to the current exploit state. \textsc{Transfer} derives triggering state from the library-side witness but does not systematically recover the client-side code that constructs or modifies the corresponding values. \textsc{Vesta} migrates concrete exploit values into reachable tests but does not determine whether intervening project logic preserves, alters, or redirects those values before the sink.

%At the same time, indiscriminately including surrounding code is ineffective. A project may contain many alternative context items including factories, implementations, accessors, transformations, and helper methods that are syntactically close to the path but irrelevant to the current exploit condition. In the long-chain cases of the benchmark, each path contains approximately 36 candidate context items on average and as many as 179. The challenge is therefore to select the code fragments that semantically govern the current exploit-relevant state, while excluding context that is merely nearby in the call graph or shares related types and interfaces.

Together, the two challenges motivate \tool: Challenge~1 identifies \emph{the exploit-relevant state required at each layer}, while Challenge~2 identifies \emph{the program context governing that state's construction and propagation}. Accordingly, \tool models layer-wise exploit-relevant states and dynamically selects the governing context.

%Together, the two challenges capture different dimensions of long-chain exploit synthesis. Challenge~1 determines \emph{what state at the current layer must satisfy the downstream exploit condition}; Challenge~2 determines \emph{which program context explains whether and how that state is constructed, transformed, propagated, or selected}. They together motivate the design of \tool, which explicitly captures layer-wise exploit-relevant states while dynamically selecting program context governing exploit-state propagation.

\section{Approach}

\begin{figure*}[htbp]
    \centering
    \includegraphics[width=\textwidth]{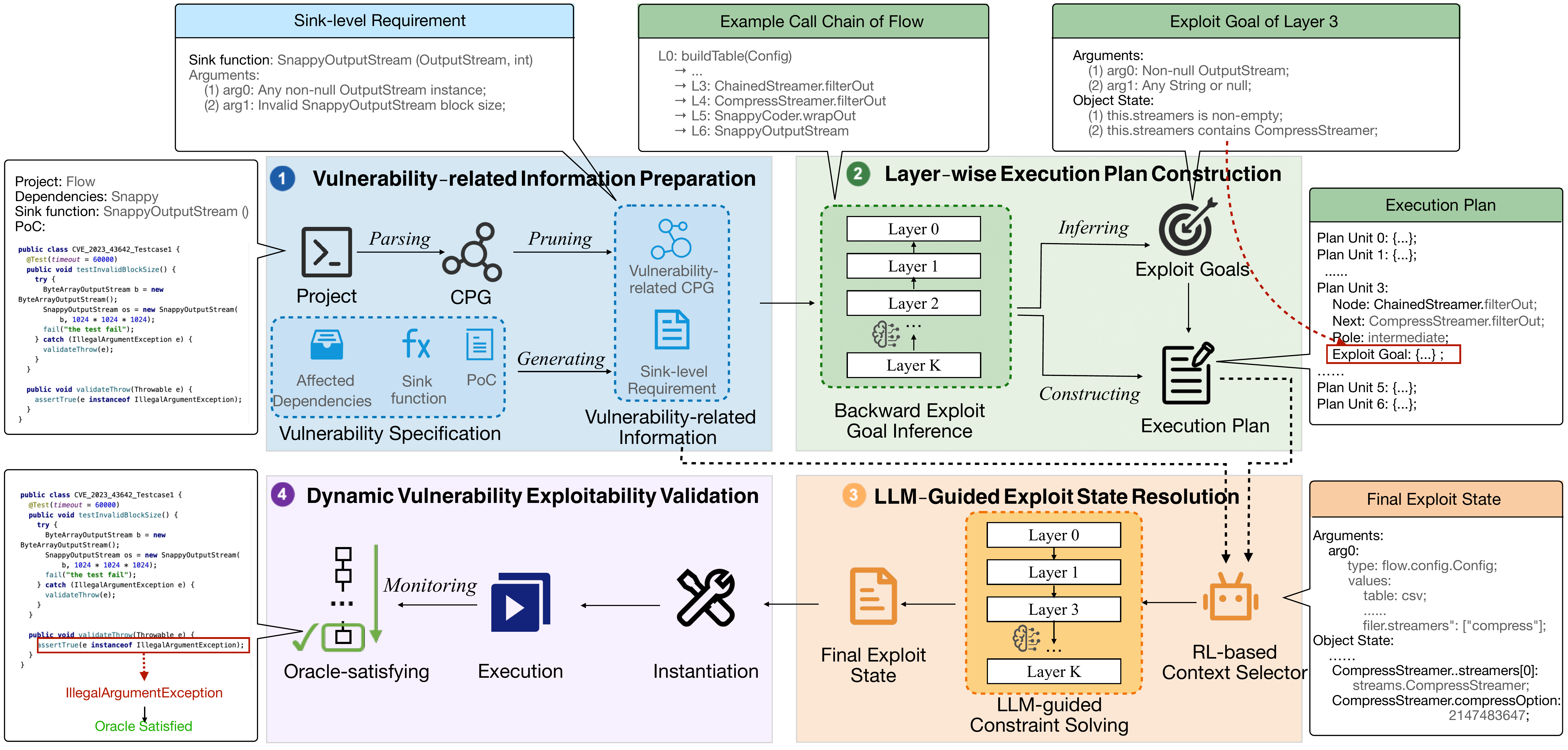} \vspace{-3mm}
    \caption{Overall framework of \tool and running example}
    \label{fig:framework}
    \vspace{-3mm}
\end{figure*}

This section presents \tool,  an LLM-guided exploit state re-expression framework with exploit-relevant context selection. As shown in Fig.~\ref{fig:framework}, we use the \texttt{flow} case from Section~\ref{sec:motivate} as a running example. \tool consists of four stages: (1) preparing vulnerability-related information, (2) inferring backward exploit goals and constructing a forward execution plan, (3) resolving exploit states through LLM-guided context selection, and (4) replaying the synthesized payload to dynamically validate exploitability.

%This section presents \tool, an LLM-guided exploit state re-expression with exploit-relevant context selection framework. As shown in Fig.~\ref{fig:framework}, we use the \texttt{flow} case introduced in Section~\ref{sec:motivate} as the running example to illustrate the workflow. \tool consists of four tightly coupled stages. First, it prepares vulnerability-related information from the project and vulnerability specification. Second, it infers exploit goals backward from the sink to the entry node and converts them into a forward execution plan that specifies what each node must resolve. Third, it performs LLM-guided exploit state resolution over the plan to produce argument and object-state bindings. Finally, it replays the synthesized payload from the project-side entry node and collects oracle-relevant evidence for the exploitability decision. 

% Algorithm~\ref{alg:vulnseer} summarizes the overall workflow. 

% \begin{algorithm}[t]
% \caption{\tool}
% \label{alg:vulnseer}
% \begin{algorithmic}[1]
% \Require Project $\mathcal{P}$, vulnerability specification $\kappa$
% \Ensure Path-payload dictionary $\mathsf{Dict}$

% \State $\mathsf{Dict} \gets \emptyset$
% \State $\mathcal{G} \gets \mathsf{BuildCPG}(\mathcal{P},D)$
% \State $\mathbb{G}_{v} \gets \mathsf{PruneRelatedCPGs}(\mathcal{G},s)$
% \State $G_L \gets \mathsf{GenerateSinkRequirement}(s,q)$

% \ForAll{$\mathcal{G}_{\rho} \in \mathbb{G}_{v}$}
%     \For{$k=L-1$ \textbf{downto} $0$}
%         \State $G_k \gets \mathsf{Infer ExploitGoal}_{\theta}(N_k,c_k,N_{k+1},G_{k+1},\mathcal{G}_{\rho})$
%     \EndFor

%     \State $\mathcal{E}_{\rho} \gets \mathsf{BuildPlan}(\rho,\langle G_0,\ldots,G_L\rangle)$
%     \State $\Sigma \gets \emptyset$

%     \For{$k=0$ \textbf{to} $L-1$}
%         \State $\widehat{C}_k \gets \mathsf{SelectContext}(\mathcal{G}_{\rho},\mathcal{E}_{\rho}[k],\Sigma)$
%         \State $\Delta_k \gets \mathsf{ResolveExploitState}_{\theta}(\mathcal{E}_{\rho}[k],\widehat{C}_k,\Sigma)$
%         \State $\Sigma \gets \mathsf{UpdateExploitState}(\Sigma,\Delta_k)$
%     \EndFor

%     \State $\pi \gets \mathsf{AssemblePayload}(\Sigma)$

%     \If{$\mathsf{Validate}(\mathcal{P},\rho,\pi,G_L)$}
%         \State $\mathsf{Dict}[\rho] \gets \pi$
%     \EndIf
% \EndFor

% \State \Return $\mathsf{Dict}$
% \end{algorithmic}
% \end{algorithm}

\subsection{Vulnerability-related Information Preparation}
\label{sec:path-recovery}
The first stage prepares the vulnerability-related information required by exploit synthesis. Given a target vulnerability, \tool takes as input a vulnerability specification $\kappa = \langle D, s, q \rangle$, where $D$ denotes the affected dependencies, $s$ denotes the vulnerable sink, and $q$ denotes a sink-level exploit sketch, i.e., a vulnerability-triggering Proof-of-Concept (PoC) at the library side.

Given a project $\mathcal{P}$ and a target vulnerability specification $\kappa$, \tool first identifies the affected dependency set
$D_{\kappa}$ specified by $\kappa$. It then retains the project code and the dependencies that may participate in propagation to the vulnerable sink, and uses Joern~\cite{joern2024workbench} to construct a code property graph (CPG) over this analysis scope. Taking the path $\rho$ starting from the project-side entry node and ending with the vulnerable sink $s$, \tool 
%prunes the CPG into a path-specific vulnerability-related CPG $\mathcal{G}_{\rho}$ by retaining the path nodes, their child nodes, and inter-routine call transitions starting from project-side entry node.
%searches the CPG for all paths that start from project-side entry node and end at $s$. For each such path $\rho$, \tool 
prunes the original CPG into a $\rho$-specific vulnerability-related CPG $\mathcal{G}_{\rho}$. 
% Specifically, \tool retains the path nodes on $\rho$, their CPG child nodes, and the call transitions between adjacent nodes on $\rho$.
%RA-Comment 1
Specifically, \tool retains the path nodes on $\rho$, their CPG child nodes, all function nodes directly connected to the call chain, and the call
transitions between adjacent nodes on $\rho$.
%Specifically, the nodes on $\rho$ are treated as parent nodes, and all their child nodes in the CPG and call transitions between adjacent routines on $\rho$ are retained. 
%The call transitions between adjacent routines on $\rho$ are also preserved.
%The collection of these path-specific CPGs is denoted as $\mathbb{G}_v$. 
The resulting CPG set $\mathbb{G}_v$ serves as the static information substrate for later stages.

In parallel, \tool converts the library-side exploit sketch $q$ into a sink-level exploit requirement. 
% Since $q$ is an executable PoC for the vulnerable dependency, an LLM extracts its vulnerability-triggering semantics, including the target sink invocation and constraints on sink arguments. For the \texttt{flow} example, the extracted requirement specifies a non-null \texttt{OutputStream} and an invalid block size.
The exploit sketch is a library-side PoC for the vulnerable dependency itself, rather than an exploit for the client project, and serves as the semantic anchor for initializing the sink-level exploit requirement. Since the exploit sketch is expressed as executable test code, \tool uses an LLM to extract its vulnerability-relevant triggering semantics, including the target sink invocation and constraints over sink arguments, thereby preserving the triggering condition at the vulnerable sink. For the \texttt{flow} example, the extracted requirement specifies a non-null \texttt{OutputStream} and an invalid block size.

\subsection{Layer-wise Execution Plan Construction}
\label{sec:backward-goal-inference}

For each path-specific vulnerability-related CPG $\mathcal{G}_{\rho} \in \mathbb{G}_v$, the second stage infers what each path node must satisfy to trigger the vulnerability.
% This process addresses \textbf{Challenge 1} by starting from the sink-level requirement and propagating it backward over the entry-to-sink path $\rho$ preserved in $\mathcal{G}_{\rho}$.
At each layer, \tool rewrites the downstream vulnerability-triggering condition into a caller-side requirement, termed an \emph{exploit goal}. We represent each exploit goal as a structured record $G=\langle E^{arg}, E^{obj}\rangle$. Here, $E^{arg}$ records constraints over arguments or local expressions used to construct the successor call, and $E^{obj}$ records
constraints over the object state that must hold before the successor call is executed, including receiver fields, nested object fields, container elements, and wrapper objects. This layer-wise backward propagation can be viewed as weak precondition inference and mirrors how a human analyst inspects an exploit path from the vulnerable sink back to the project-side entry node. As shown in Fig.~\ref{fig:framework}, in the \texttt{flow} case, the goal for \texttt{ChainedStreamer.filterOut} requires a non-null output stream at the argument level and a receiver state where \texttt{this.streamers} is non-empty and contains a \texttt{CompressStreamer}.

\subsubsection{Backward Exploit Goal Inference}

Starting from the sink-level requirement, \tool performs backward propagation on each $\mathcal{G}_{\rho}$.
Let $G_k$ denote the exploit goal that has been inferred for the current callee node $N_k$.
For the backward step from $N_k$ to its caller $N_{k-1}$, \tool analyzes the call edge $c_{k-1}$ and infers a caller-side exploit goal $G_{k-1}$ that enables $N_{k-1}$ to invoke $N_k$ in a state satisfying $G_k$. 
To support this inference, \tool retrieves goal-relevant context from $\mathcal{G}_{\rho}$, including
the signatures of the caller ($N_{k-1}$) and callee ($N_k$), the formal parameters of the caller $N_{k-1}$, the pruned source code of $N_{k-1}$, helper methods invoked by $N_{k-1}$ that participate in call construction, and class-level context of the caller-side receiver, including fields and constructors.
The pruned source code of $N_{k-1}$ is constructed by locating the successor call site $c_{k-1}$ in $N_{k-1}$ and retaining the minimal enclosing CPG subtree needed to preserve the local call construction.
Given this context, the LLM infers a compact exploit goal $G_{k-1}$ over the state visible to the caller node $N_{k-1}$ rather than a concrete payload. 
%RA-Comment 1
Repeating this process from $N_L$ to $N_0$ produces a sequence of layer-wise exploit goals.

%The LLM is prompted to infer a sufficient caller-side exploit goal under which executing the relevant code fragment of $N_{k-1}$ can produce a call to $N_k$ satisfying the current callee-side exploit goal $G_k$. Its output is not a concrete payload, but a compact caller-side exploit goal $G_{k-1}$ which is expressed over the state visible to the caller node $N_{k-1}$. 
%Repeating this process from $N_L$ to $N_0$ yields a complete sequence of layer-wise exploit goals.

\subsubsection{Execution Plan Construction}
\label{sec:exe_plan}
Backward inference produces a sequence of exploit goals from the sink to the entry, specifying the caller-side conditions required at each layer. Yet, they do not provide an operational order for synthesizing an executable entry-side payload. Since exploit synthesis proceeds in the opposite direction, \tool reorganizes these goals into an entry-to-sink execution plan $\mathcal{E}_{\rho}$ that converts discrete backward goals into the forward solving order. As shown in Fig.~\ref{fig:framework}, each plan unit corresponds to a node in $\mathcal{G}_{\rho}$ and records the node with its pruned source code, its role (entry, intermediate, or sink), its successor call (if any), and the local exploit goal to satisfy, making each step a local exploit-state transition.

%Backward inference yields a sequence of layer-wise exploit goals in the sink-to-entry direction. These goals specify what must hold on the caller side at each layer to make the downstream requirement satisfiable, but they do not provide an operational order for synthesizing an executable entry-side payload. A real exploit executes in the opposite direction, from the project-side entry node to the vulnerable sink. \tool therefore reorganizes the backward-derived goals into an entry-to-sink execution plan $\mathcal{E}_{\rho}$ for the current $\mathcal{G}_{\rho}$. The purpose of the execution plan is to convert discrete backward goals into a forward solving schedule. 
%As shown in Fig.~\ref{fig:framework}, each plan unit corresponds to one node in $\mathcal{G}_{\rho}$ and records the current node, its role (entry, intermediate, or sink) in the path, the successor call node if any, and the local exploit goal to be satisfied at that node. This makes each solving step a local exploit-state transition over the vulnerability-related code.

\subsection{LLM-Guided Exploit State Resolution}
\label{sec:state-resolution}

In the third stage, \tool performs LLM-guided exploit state resolution over the execution plan generated in Section~\ref{sec:exe_plan}. Starting from the project-side entry node, \tool traverses $\mathcal{E}_{\rho}$ and incrementally constructs the exploit state $\Sigma$. 
At each layer $k$, the current plan unit $\mathcal{E}_{\rho}[k]$ provides the path node, its pruned source code, and the local exploit goal to be resolved. The RL-based context selector then retrieves additional exploit-relevant context $\widehat{C}_k$ from $\mathcal{G}_{\rho}$ according to the current plan unit and the accumulated exploit state.
Given the selected context $\widehat{C}_k$, the LLM resolver takes $\mathcal{E}_{\rho}[k]$ and $\Sigma$ as input, produces a local state update $\Delta_k$, and applies it to $\Sigma$ by extending the argument bindings and object-state bindings needed to satisfy the current exploit goal and preserve the downstream vulnerability-triggering condition. For example, as shown in Fig.~\ref{fig:framework}, when resolving the layer from \texttt{ChainedStreamer.filterOut} to \texttt{CompressStreamer.filterOut}, $\widehat{C}_k$ supplies off-chain evidence for constructing \texttt{this.streamers}, such as the streamer registry and the factory logic reading \texttt{filer.streamers} from \texttt{Config}. Given $\mathcal{E}_{\rho}[k]$, $\Sigma$, and $\widehat{C}_k$, the resolver emits an update $\Delta_k$, e.g., \texttt{out != null} and \texttt{this.streamers instanceof CompressStreamer}. This update is merged into $\Sigma$ for later layers.

\subsubsection{RL-based exploit-relevant context selection}
% As discussed in Challenge~2, the key difficulty is not simply collecting more surrounding code, but selecting the exploit-relevant construction context from noisy project code. Even though $\mathcal{G}_{\rho}$ has already pruned the analysis scope around the target vulnerability, it may still contain distracting or low-value fragments, especially around critical branch points where excessive context can cause reasoning drift or prompt explosion. 
% To address this issue, \tool employs a learned context selector to choose a compact and exploit-relevant subset before invoking the exploit-state resolver.

The RL-based context selector is trained on the selector training split
(described in Section~\ref{sec:rl-dataset}). 
For each layer, \tool queries the path-specific vulnerability-related CPG $\mathcal{G}_{\rho}$ and obtains a set of additional candidate context items beyond the pruned source code already carried by $\mathcal{E}_{\rho}[k]$. These items include call-site evidence, object-state definitions, branch predicates, and helper functions that may affect the current exploit goal resolution. 
The selector is trained as a sequential binary policy that decides whether to include each candidate as an extra input to the exploit-state resolver based on the current plan unit, accumulated exploit state, the candidate item, and the previously selected context $\widehat{C}_k$. The action space is binary, which means the selector either includes or excludes the candidate item.
Included items and corresponding source code are materialized into the selected context $\widehat{C}_k$.
%When the item is backed by source code in $\mathcal{G}_{\rho}$, \tool includes the corresponding code slice.
When source code is unavailable (e.g., unresolved external or library calls), \tool substitutes a conservative semantic summary derived from the call signature and observed usage.

%We train the context selector as a sequential binary policy over the candidate context items. The environment presents the items one by one. At each decision step, the selector observes the current plan unit, a compact summary of the accumulated exploit state, the current candidate item, and the context already selected in the current layer. 

%After all candidates are processed, the selected context is passed to the exploit-state resolver.
%If an item is included, \tool materializes it into the selected context. When the item is backed by source code in $\mathcal{G}_{\rho}$, \tool includes the corresponding code slice. When exact source code is unavailable, such as for an unresolved external or library call, \tool falls back to a conservative semantic summary grounded in the call signature and observed usage, without inventing its internal implementation. Excluded items are not passed to the resolver.
%After all context items in the current layer are processed, the selected context is passed to the exploit-state resolver.

The terminal validation signal directly reflects exploit quality, but it is sparse and delayed. The selector observes it only after dynamic validation is completed, making it difficult to determine which individual context item helped preserve the exploit condition. We therefore introduce a local reward to shape each intermediate context-selection decision: $r^{\mathsf{local}}_{k,i}
=
\alpha_{\mathsf{1}} r_{\mathsf{sil}}
+
\alpha_{\mathsf{2}} r_{\mathsf{type}}
-
\alpha_{\mathsf{3}} r_{\mathsf{len}}$,
where $x_{k,i}$ denotes the $i$-th context item considered at layer $k$.
Here, $r_{\mathsf{sil}}$ distinguishes whether the context item is backed by source code. We set $r_{\mathsf{sil}}=1$ when the candidate is supported by a code slice, and $r_{\mathsf{sil}}=0$ when the slicer fails and a conservative semantic summary is used. 
The type prior $r_{\mathsf{type}}$ is assigned by the role of the context item. Exploit-state construction context, including fields, constructors, factory logic, configuration accessors, container updates, and wrapper relations, receives $r_{\mathsf{type}}=1$. Exploit-state propagation context, including successor-call arguments, receiver expressions, branch predicates, and dynamic-dispatch targets, receives $r_{\mathsf{type}}=0.5$. Low-relevance context, including logging, comments, formatting utilities, unrelated accessors, and formatting utilities, receives $r_{\mathsf{type}}=0$. The cost-related term $r_{\mathsf{len}}(x_{k,i})=\min(\mathsf{Tok}(x_{k,i})/T_{\max},1)$ penalizes the normalized token cost of $x_{k,i}$, where $\mathsf{Tok}(x_{k,i})$ is its token length and $T_{\max}$ is the context-item token budget.

The coefficients $\alpha_{\cdot}$ control the relative contribution of each local feedback term. We set $(\alpha_{\mathsf{1}}, \alpha_{\mathsf{2}}, \alpha_{\mathsf{3}})$ to $(0.5, 0.4, 0.1)$ in all experiments, giving priority to source-code support and context-item type while using a smaller penalty for context length. The terminal reward is derived from the dynamic exploitability validation (Sec.\ref{sec:dynamic-validation}) result of the final payload. 

The final reward assigned to a decision is $R_{k,i}
=
(1-\lambda)\,r^{\mathsf{local}}_{k,i}
+
\lambda\,
r^{\mathsf{terminal}}$,
We set $\lambda=0.3$ in all experiments. This design uses the local reward to provide dense feedback for intermediate selection decisions, while the terminal reward ties the selector to the final exploitability outcome and discourages choices that look locally useful but do not help the downstream resolver synthesize a valid payload.

The selector is implemented with Qwen3-8B and optimized using PPO with LoRA adaptation. At each decision step, it predicts a binary action indicating whether to include the current context item. To stabilize training, we initialize the policy with LoRA-based supervised fine-tuning on manually annotated selector decisions from the reserved training split. 
Two Java security experts independently label each context item according to whether it helps construct, store, or propagate the current exploit goal. Items that support exploit-state construction or propagation are labeled positive, while items that mainly introduce prompt noise, including logging, formatting utilities, unrelated accessors, or comments, are labeled negative. The two annotators achieve a Cohen's kappa of 0.76.
Disagreements are resolved through discussion.
%Two Java security experts independently labeled whether each context item contributes to exploit-state construction or propagation, achieving a Cohen's Kappa of 0.76. 
These labels are used only for initialization; the final policy is optimized by PPO with the reward above.

%The selector policy is implemented with Qwen3-8B as the base decoder-only model. We adapt it using LoRA and attach a value head for PPO training. At each decision step, the model is prompted with the rendered selector state and trained to emit a single token, either ``0'' or ``1'', indicating whether to include or exclude the current context item.
%To stabilize PPO training, we warm-start the policy with LoRA-based supervised fine-tuning and manually labeled selector decisions from the reserved selector-training split. No evaluation-set call chain is used. Two Java security experts independently label each context item according to whether it helps construct, store, or propagate the current exploit goal. Items that support exploit-state construction or propagation are labeled positive, while items that mainly introduce prompt noise, such as logging, formatting utilities, unrelated accessors, or comments, are labeled negative. The two annotators achieve a Cohen's Kappa of 0.76.
%Disagreements are resolved through discussion. These labels are used only for policy initialization, while the final selector is optimized by PPO with the reward signals defined above.

\subsubsection{LLM-Guided Constraint Solving}

At layer $k$, the context selector retrieves goal-relevant context $\widehat{C}_k$ from $\mathcal{G}_{\rho}$ according to the current plan unit $\mathcal{E}_{\rho}[k]$, which includes current node, its successor node, and the local exploit goal $G_k$.
Then, the pruned source code carried by $\mathcal{E}_{\rho}[k]$ and the selected additional context $\widehat{C}_k$ are rendered together as a Java sub-program $S_k$ for resolving the current layer.
Let the accumulated exploit state be $\Sigma_k=\langle B_k^{arg},B_k^{obj}\rangle$, where $B_k^{arg}$ records argument and $B_k^{obj}$ records receiver fields, nested object fields, container elements, and wrapper object state bindings. The resolver searches for a local update $\Delta_k$ such that \(\Sigma_k \oplus \Delta_k \models G_k\) and the Hoare triple~\cite{hoare1969axiomatic} \(\{\Sigma_k \oplus \Delta_k\}\ S_k\ \{G_{k+1}\}\) holds. The resulting $\Delta_k$ is merged into $\Sigma_k$, enabling incremental exploit-state construction across layers. 
As shown in Fig.~\ref{fig:framework}, in the \textsc{Flow} case, the final exploit state binds the entry argument to a \texttt{Config} object whose fields enable the compress streamer path, e.g., setting \texttt{filer.streamers} to include \texttt{"compress"}, and records downstream object-state bindings such as \texttt{CompressStreamer.compressOption = 2147483647} to satisfy the sink-side block-size requirement.

\subsection{Dynamic Exploitability Validation}
\label{sec:dynamic-validation}

The final stage validates whether the inferred exploit state is executable. Given the final exploit state $\Sigma=\langle B^{arg},B^{obj}\rangle$, \tool assembles an executable payload $\pi=\langle A,F\rangle$.
The entry-argument map $A$ is extracted from $B^{arg}$ and binds each entry parameter to a concrete value. The object-state map $F$ is extracted from $B^{obj}$ and records receiver fields, nested fields, container elements, wrapper objects, and static state that must be instantiated before replay.
\tool then performs payload instantiation in an instrumented runtime. For entry arguments, it converts the values in $A$ to the declared argument types. For receiver and object states in $F$, it instantiates the required objects using accessible constructors when possible. 
% If no suitable constructor is available, it allocates the object without running its constructor and then initializes the required state.
%RB-Comment 4：Without an Available Constructor
If no suitable constructor is available, \tool uses \texttt{Unsafe.allocateInstance} from the provided JDK to allocate an instance of the required runtime class without executing its constructor. It then restores the object's fields according to the recorded replay state before invoking the target method.
If the required state cannot be instantiated, the replay fails before oracle evaluation.

%The final stage validates whether the final exploit state can be realized under concrete execution. Given the final exploit state $\Sigma=\langle B^{arg},B^{obj}\rangle$, \tool first assembles an executable payload $\pi=\langle A,F\rangle$. The entry-argument map $A$ is extracted from $B^{arg}$ and binds each entry parameter to a concrete value. The object-state map $F$ is extracted from $B^{obj}$ and records receiver fields, nested fields, container elements, wrapper objects, and static state that must be instantiated before replay. \tool then performs payload instantiation in an instrumented runtime. For entry arguments, it converts the values in $A$ to the declared argument types. For receiver and object states, it instantiates the required objects using accessible constructors when possible. If no suitable constructor is available, it allocates the object without running its constructor and then initializes the required state. If the required state cannot be instantiated, the replay fails before oracle evaluation.

After instantiation, \tool executes the designated project-side entry node and monitors runtime evidence using an ASM-instrumented class loader~\cite{asm2024}, including invoked method signatures, sink arguments, receiver objects, return values, and thrown exceptions. 
The validation result is determined by an \textbf{Oracle-satisfying} check, which examines whether the observed runtime evidence matches the expected vulnerable behavior specified by the sink-level requirement and the vulnerability oracle. In particular, \tool uses the monitored trace to identify oracle-relevant sink-side evidence, such as the observed sink arguments and thrown exception. As shown in Fig.~\ref{fig:framework}, in the \textsc{Flow} case, the oracle is satisfied when the monitored execution observes a non-null \texttt{OutputStream}, an invalid block size, and the expected \texttt{IllegalArgumentException}.

\section{Experiment Setup}
We use the same implementation and configuration described in Section~\ref{sec:motivate}. 

We design the following research questions.

\begin{itemize}
    \item \textbf{RQ1:}
    What is the overall performance of  VulnSEER in generating verified exploits for vulnerable dependency call chains compared with baselines?
    % We compare VulnSEER with state-of-the-art baselines on the expanded ground-truth benchmark.

    \item \textbf{RQ2:}
    How much does each design choice contribute to VulnSEER's effectiveness?
    % We evaluate several variants of VulnSEER to quantify the impact of its key components.

    \item \textbf{RQ3:}
    What are the execution time and token costs of VulnSEER?
    % We measure the wall-clock time and the LLM-related cost required to generate verified exploits.
\end{itemize}

\subsubsection{Benchmark Construction}
\label{sec:rl-dataset}
We build our evaluation benchmark on top of the \textsc{Magneto} benchmark, which contains 32 real-world Java vulnerabilities, 45 upstream vulnerable functions, 21 vulnerable dependency libraries, 49 client projects, and 182 ground-truth exploitable call chains. To better evaluate exploit generation under complex dependency scenarios, we expand this ground-truth corpus with additional client-side paths whose vulnerability-triggering conditions must be propagated across multiple caller-callee boundaries.
%RA-Comment 4：Benchmark Expansion
The added chains come from the same vulnerability-client pairs in the original benchmark and reuse the corresponding vulnerability-specific oracles.
Specifically, for each vulnerability-client pair, we build an inter-procedural call graph over the client project and its vulnerable dependencies, locate the vulnerability-specific sink in the dependency library, and search for statically reachable paths from public client-side entry functions to the sink.

Two Java experts then spent two months manually inspecting and validating the candidate chains. A chain is included in the ground truth only if both experts confirm that it is exploitable and can be validated by a vulnerability-specific oracle. 
% Answer to RC_Q1&comment1: JUnit/Oracle Test
The added chains reuse the vulnerability-specific JUnit oracle of the corresponding vulnerability without modification. The benchmark expansion therefore preserves the original vulnerability set and success criterion.
Disagreements were resolved through joint discussion and re-execution of the corresponding oracle tests. After this process, the expanded corpus contains 319 ground-truth exploitable call chains, including 89 direct calls, 204 short chains, and 26 long chains.

\subsubsection{Evaluation and Training Split}
Among the 319 exploitable call chains, we reserve 50 chains from a disjoint set of vulnerabilities as the selector-training set, including 11 direct calls, 31 short chains, and 8 long chains. These chains are used only to train the RL-based context selector and its alternative selector variants, and none of them shares the same CVE with the evaluation set. The remaining 269 chains constitute the evaluation benchmark used for comparing \tool with baselines and ablation variants, including 78 direct calls, 173 short chains, and 18 long chains.

\subsubsection{Baselines}
Following the baseline selection rationale in Section~\ref{sec:motivate}, we compare \tool with \textsc{Magneto}, \textsc{Vesta}, \textsc{Transfer}, and \textsc{VulEUT}~\cite{gao2025vulnerability}.
%RB_Q2&comment7
VulEUT analyzes reachable paths from client-side code to vulnerable library functions and identifies how vulnerability-triggering parameters are transferred across the path. It classifies parameter transfer into direct propagation, type conversion, value change, and non-propagation, and generates vulnerability tests based on the inferred path and transfer information. 
In this section, we use these tools as full experimental baselines on the expanded benchmark.
For all LLM-based tools, we use the same backend set (GPT-5.2, GPT-4o, Claude-Sonnet-4.6, DeepSeek-V3.2, and Qwen3.5-397B-A17B), API snapshot,
temperature, maximum output length, API-retry budget, and timeout. 
%RA-Q1 & Comment 2
The API-retry budget covers only individual LLM calls that fail or return malformed responses. Each tool performs one end-to-end synthesis run per call chain per experimental run.
All experiments are executed on the same machine under the same parallelism setting. For \textsc{Magneto}, we use the authors' released implementation and adapt only the model invocation interface to ensure
the same backend configuration.

\subsubsection{Evaluation Metrics}
We use the same chain-level success rate (SR) introduced in
Section~\ref{sec:motivate} as the primary evaluation metric for RQ1 and RQ2. For RQ3, we report execution time efficiency and token cost. 
Execution time is measured as end-to-end wall-clock time from receiving a project to generating either a validated exploit or a failure decision. Token cost includes the input and output tokens consumed by all LLM invocations in the framework.

\section{Evaluation}
\subsection{RQ1: Overall Effectiveness}

\begin{table*}[t]
\centering
\caption{Effectiveness comparison under different solver models and baselines.}
\label{tab:rq1-effectiveness}

\footnotesize
\setlength{\tabcolsep}{1.8pt}
\renewcommand{\arraystretch}{1.15}

\resizebox{\textwidth}{!}{%
\begin{tabular}{@{}lcc*{13}{c}@{}}
\toprule
\multirow{3}{*}{\makecell{\textbf{Category}}}
& \multirow{3}{*}{\makecell{\textbf{Length}}}
& \multirow{3}{*}{\makecell{\textbf{Ground}\\\textbf{Truth}}}
& \multicolumn{5}{c}{\textbf{\tool}}
& \multicolumn{5}{c}{\textbf{\textsc{Magneto}}}
& \multicolumn{3}{c}{\textbf{Other Baselines}} \\
\cmidrule(lr){4-8}
\cmidrule(lr){9-13}
\cmidrule(lr){14-16}
&
&
& \textbf{GPT-5.2}
& \textbf{GPT-4o}
& \textbf{Claude}
& \textbf{DeepSeek}
& \textbf{Qwen}
& \textbf{GPT-5.2}
& \textbf{GPT-4o}
& \textbf{Claude}
& \textbf{DeepSeek}
& \textbf{Qwen}
& \textbf{\textsc{Vesta}}
& \textbf{\textsc{Transfer}}
& \textbf{\textsc{VulEUT}} \\
\cmidrule(lr){4-16}
&
&
& \textbf{SR}
& \textbf{SR}
& \textbf{SR}
& \textbf{SR}
& \textbf{SR}
& \textbf{SR}
& \textbf{SR}
& \textbf{SR}
& \textbf{SR}
& \textbf{SR}
& \textbf{SR}
& \textbf{SR}
& \textbf{SR} \\
\midrule

Direct
& 1
& 78
& \makecell{\textbf{98.72\%}\\[-1mm]{\scriptsize ($\pm$0.00)}}
& \makecell{\textbf{98.72\%}\\[-1mm]{\scriptsize ($\pm$0.00)}}
& \makecell{\textbf{98.72\%}\\[-1mm]{\scriptsize ($\pm$0.00)}}
& \makecell{\textbf{98.72\%}\\[-1mm]{\scriptsize ($\pm$0.00)}}
& \makecell{\textbf{98.72\%}\\[-1mm]{\scriptsize ($\pm$0.00)}}
& \textbf{98.72\%}
& \textbf{98.72\%}
& \textbf{98.72\%}
& \textbf{98.72\%}
& \textbf{98.72\%}
& 55.13\%
& 57.69\%
& 35.90\% \\

\midrule
\multirow{4}{*}{Short}
& 2
& 54
& \makecell{\textbf{95.68\%}\\[-1mm]{\scriptsize ($\pm$1.23)}}
& \makecell{88.27\%\\[-1mm]{\scriptsize ($\pm$1.23)}}
& \makecell{\textbf{95.68\%}\\[-1mm]{\scriptsize ($\pm$1.23)}}
& \makecell{92.59\%\\[-1mm]{\scriptsize ($\pm$1.85)}}
& \makecell{87.04\%\\[-1mm]{\scriptsize ($\pm$1.85)}}
& 51.85\%
& 38.89\%
& 51.85\%
& 38.89\%
& 37.04\%
& 1.85\%
& 5.56\%
& 16.67\% \\

& 3
& 51
& \makecell{\textbf{90.20\%}\\[-1mm]{\scriptsize ($\pm$1.96)}}
& \makecell{71.90\%\\[-1mm]{\scriptsize ($\pm$1.31)}}
& \makecell{88.89\%\\[-1mm]{\scriptsize ($\pm$1.31)}}
& \makecell{82.35\%\\[-1mm]{\scriptsize ($\pm$1.96)}}
& \makecell{71.90\%\\[-1mm]{\scriptsize ($\pm$1.31)}}
& 49.02\%
& 43.14\%
& 47.06\%
& 47.06\%
& 43.14\%
& 0.00\%
& 9.80\%
& 21.57\% \\

& 4
& 43
& \makecell{74.42\%\\[-1mm]{\scriptsize ($\pm$0.00)}}
& \makecell{59.69\%\\[-1mm]{\scriptsize ($\pm$1.55)}}
& \makecell{\textbf{76.74\%}\\[-1mm]{\scriptsize ($\pm$0.00)}}
& \makecell{69.77\%\\[-1mm]{\scriptsize ($\pm$0.00)}}
& \makecell{55.81\%\\[-1mm]{\scriptsize ($\pm$2.33)}}
& 44.19\%
& 34.88\%
& 41.86\%
& 44.19\%
& 27.91\%
& 0.00\%
& 0.00\%
& 0.00\% \\

& 5
& 25
& \makecell{64.00\%\\[-1mm]{\scriptsize ($\pm$0.00)}}
& \makecell{64.00\%\\[-1mm]{\scriptsize ($\pm$0.00)}}
& \makecell{\textbf{68.00\%}\\[-1mm]{\scriptsize ($\pm$0.00)}}
& \makecell{64.00\%\\[-1mm]{\scriptsize ($\pm$0.00)}}
& \makecell{52.00\%\\[-1mm]{\scriptsize ($\pm$0.00)}}
& 32.00\%
& 32.00\%
& 40.00\%
& 32.00\%
& 32.00\%
& 0.00\%
& 0.00\%
& 0.00\% \\

\midrule
\multirow{3}{*}{Long}
& 6
& 9
& \makecell{\textbf{74.07\%}\\[-1mm]{\scriptsize ($\pm$7.41)}}
& \makecell{62.96\%\\[-1mm]{\scriptsize ($\pm$7.41)}}
& \makecell{66.67\%\\[-1mm]{\scriptsize ($\pm$0.00)}}
& \makecell{62.96\%\\[-1mm]{\scriptsize ($\pm$7.41)}}
& \makecell{62.96\%\\[-1mm]{\scriptsize ($\pm$7.41)}}
& 11.11\%
& 11.11\%
& 11.11\%
& 11.11\%
& 11.11\%
& 0.00\%
& 0.00\%
& 0.00\% \\

& 7
& 6
& \makecell{\textbf{50.00\%}\\[-1mm]{\scriptsize ($\pm$0.00)}}
& \makecell{16.67\%\\[-1mm]{\scriptsize ($\pm$0.00)}}
& \makecell{33.33\%\\[-1mm]{\scriptsize ($\pm$0.00)}}
& \makecell{33.33\%\\[-1mm]{\scriptsize ($\pm$0.00)}}
& \makecell{16.67\%\\[-1mm]{\scriptsize ($\pm$0.00)}}
& 0.00\%
& 0.00\%
& 0.00\%
& 0.00\%
& 0.00\%
& 0.00\%
& 0.00\%
& 0.00\% \\

& 8
& 3
& \makecell{\textbf{66.67\%}\\[-1mm]{\scriptsize ($\pm$0.00)}}
& \makecell{33.33\%\\[-1mm]{\scriptsize ($\pm$0.00)}}
& \makecell{44.44\%\\[-1mm]{\scriptsize ($\pm$22.22)}}
& \makecell{\textbf{66.67\%}\\[-1mm]{\scriptsize ($\pm$0.00)}}
& \makecell{33.33\%\\[-1mm]{\scriptsize ($\pm$0.00)}}
& 0.00\%
& 0.00\%
& 0.00\%
& 0.00\%
& 0.00\%
& 0.00\%
& 0.00\%
& 0.00\% \\

\midrule
Total
& --
& 269
& \makecell{\textbf{87.11\%}\\[-1mm]{\scriptsize ($\pm$0.62)}}
& \makecell{78.31\%\\[-1mm]{\scriptsize ($\pm$0.50)}}
& \makecell{86.74\%\\[-1mm]{\scriptsize ($\pm$0.25)}}
& \makecell{83.52\%\\[-1mm]{\scriptsize ($\pm$0.99)}}
& \makecell{76.33\%\\[-1mm]{\scriptsize ($\pm$1.24)}}
& \makecell{58.74\%\\[-1mm]{\scriptsize \textcolor{red}{$\uparrow$28.37\%}}}
& \makecell{53.53\%\\[-1mm]{\scriptsize \textcolor{red}{$\uparrow$24.78\%}}}
& \makecell{58.74\%\\[-1mm]{\scriptsize \textcolor{red}{$\uparrow$28.00\%}}}
& \makecell{55.76\%\\[-1mm]{\scriptsize \textcolor{red}{$\uparrow$27.76\%}}}
& \makecell{52.04\%\\[-1mm]{\scriptsize \textcolor{red}{$\uparrow$24.29\%}}}
& \makecell{16.36\%\\[-1mm]{\scriptsize \textcolor{red}{$\uparrow$70.75\%}}}
& \makecell{19.70\%\\[-1mm]{\scriptsize \textcolor{red}{$\uparrow$67.41\%}}}
& \makecell{17.84\%\\[-1mm]{\scriptsize \textcolor{red}{$\uparrow$69.27\%}}} \\

\bottomrule
\end{tabular}%
}
\vspace{-2mm}
\end{table*}

\subsubsection{Strategy}
This RQ evaluates the effectiveness of \tool in generating verified exploits. 
Following Section~\ref{sec:motivate}, we divide call chains into direct calls (length 1), short chains (length 2-5), and long chains (length $\geq$ 6), allowing us to examine whether \tool remains effective as the exploit condition propagates through more caller-callee transitions.
We compare \tool with \textsc{Transfer}, \textsc{Vesta}, \textsc{Magneto} and \textsc{VulEUT}. 
For the LLM-based methods, namely \tool and \textsc{Magneto}, we use the same set of backbone models, including GPT-5.2, GPT-4o, Claude-Sonnet-4.6, DeepSeek-V3.2, and Qwen3.5-397B-A17B.
%RB_Q2&comment7
\textsc{Transfer}, \textsc{Vesta}, and
\textsc{VulEUT} are evaluated using their original implementations and default configurations.
%RB_comment6: Repeated Runs
To evaluate the stability of \tool under the stochasticity inherent in LLM-based generation, we execute each of its five LLM configurations three times on the same set of 269 call chains. We report the arithmetic mean success rate across the three executions, with the $\pm$ value in parentheses indicating the maximum absolute deviation of an individual execution from the mean, measured in percentage points. All results are reported for each call-chain length and aggregated across direct calls, short chains, and long chains.
% We report the success rate for each call-chain length, aggregated over direct calls, short chains, and long chains.

\subsubsection{Results}
Table~\ref{tab:rq1-effectiveness} reports the effectiveness comparison across all baselines and solver models. 
% Under GPT-5.2, \tool solves 236 out of 269 chains, achieving an overall success rate of 87.73\%, while \textsc{Magneto}, \textsc{Vesta}, and \textsc{Transfer} achieve 58.74\%, 16.36\%, and 19.70\%, respectively. 
%RB_Q2&comment7 and RB_comment6: Repeated Runs
Under GPT-5.2, \tool achieves a mean overall success rate of 87.11\% ($\pm$0.62\%), while \textsc{Magneto}, \textsc{Vesta}, \textsc{Transfer}, and \textsc{VulEUT} achieve 58.74\%, 16.36\%, 19.70\%, and 17.84\%, respectively. This advantage is consistent across LLM backends: \tool achieves mean success rates of 87.11\%, 78.31\%, 86.74\%, 83.52\%, and 76.33\% under GPT-5.2, GPT-4o, Claude, DeepSeek, and Qwen, 
compared with only 58.74\%, 53.53\%, 58.74\%, 55.76\%, and 52.04\% for the corresponding \textsc{Magneto} variants.
This corresponds to an improvement of 24.29--28.37 percentage points.

%RB_comment6: Repeated Runs
Across the three repeated runs, the maximum deviation from the mean overall success rate is at most 1.24 percentage points across all solver models. The deviations are particularly small for GPT-5.2 and Claude, at 0.62 and 0.25 percentage points, respectively, while Qwen exhibits the largest variation of 1.24 percentage points. These results indicate that \tool remains stable across repeated runs.

For direct calls, both \tool and \textsc{Magneto} are already highly effective, solving 77 out of 78 cases under all LLM backends, so \tool's main advantage comes from cases requiring reasoning over additional caller-side context. 
%RB_comment6: Repeated Runs
For short chains, \tool achieves a mean success rate of 84.20\% under GPT-5.2, compared with 46.24\% for \textsc{Magneto}. Across the other solver models, the mean short-chain success rate of \tool ranges from 69.75\% to 84.97\%, whereas the corresponding \textsc{Magneto} results range from 35.84\% to 46.24\%.
%RA-Q1 & Comment 2
\textsc{Vesta}, \textsc{Transfer}, and \textsc{VulEUT} only solve 1, 8, and 20 short-chain cases, respectively.
%RB_comment6: Repeated Runs
The gap widens further on long chains: among the 18 ground-truth chains with length at least 6, \tool achieves a mean success rate of 64.81\% under GPT-5.2, compared with 5.56\% for \textsc{Magneto}. Across the other solver models, the mean long-chain success rate of \tool ranges from 42.59\% to 53.70\%, whereas \textsc{Magneto} remains at 5.56\%. 

%RA-Q1 & Comment 2
\textsc{Vesta}, \textsc{Transfer}, and \textsc{VulEUT} fail to solve any long-chain case.

Qualitatively, this improvement stems from \tool's two-stage reasoning over multi-layer call chains. 
The \texttt{flow} case illustrates this distinction: reaching the vulnerable constructor requires more than forwarding an oversized block-size argument backward through the chain, as the downstream requirement is successively represented by a receiver field, a dynamic coder target, container membership, and ultimately configuration entries driving factory construction. 
Baselines transfer exploit values or local callee states without maintaining correspondence between these changing representations or retrieving the off-chain code that realizes them. 
\tool instead records the required state at each layer as an exploit goal and resolves it using selected factory, registry, constructor, and field-initialization context, explaining why its advantage over baselines grows with chain depth.

%RA-Q1 & Comment 2
To examine the remaining limitations, we analyze failures under GPT-5.2, the best-performing solver model for \tool in RQ1. In the reported run, \tool fails on 33 of the 269 evaluation chains. Manual inspection identifies incomplete cross-layer exploit-state re-expression in seven cases and missing off-chain propagation context in three. Most remaining failures involve uncontrollable third-party API data or runtime issues. The unsolved \href{https://github.com/bit15k/Wechat-SSM}{\textcolor{blue}{\texttt{wechat-ssm}}}
case (\href{https://nvd.nist.gov/vuln/detail/CVE-2022-25845}{\textcolor{blue}{CVE-2022-25845}}) illustrates the limitation imposed by external API data; neither \tool nor any baseline succeeds. Here, the obstacle is the source of the sink input rather than cross-layer state re-expression. The chain reaches Fastjson through
\texttt{getJsapiSigner}, \texttt{getJsapiTicket}, and \texttt{getAccessToken}. However, the parsed value, \texttt{responseStr}, comes from an HTTP response
returned by the WeChat server. The user-controlled \texttt{url} parameter only affects JSAPI signing and does not flow into \texttt{responseStr}. Triggering
the vulnerability therefore requires control over the external response, which falls outside our entry-payload threat model.

% We also examine an unsolved \href{https://github.com/bit15k/Wechat-SSM}{\textcolor{blue}{\texttt{wechat-ssm}}} case (\href{https://nvd.nist.gov/vuln/detail/CVE-2022-25845}{\textcolor{blue}{CVE-2022-25845}}), where neither \tool nor any baseline succeeds. 
% Unlike the \texttt{flow} case, the failure is not caused by losing object-state constraints across wrappers: the chain reaches Fastjson through \texttt{getJsapiSigner}, \texttt{getJsapiTicket}, and \texttt{getAccessToken}, but the value parsed by Fastjson, \texttt{responseStr}, originates from an HTTP response returned by the WeChat server rather than from the user-controlled \texttt{url} parameter, which only affects JSAPI signing. 
% Satisfying the sink would require controlling this external server response rather than constructing a different entry-side payload, a source-sink connection absent under our entry-payload threat model.

\begin{tcolorbox}
\textbf{Answer to RQ1:} 
\tool achieves the highest overall effectiveness across all evaluated solver models.
Its advantage is evident in short-chain cases and becomes more substantial in long-chain cases, demonstrating stronger robustness as call-chain complexity increases.
Its stable performance across repeated runs further supports the robustness of \tool.
\end{tcolorbox}

\subsection{RQ2: Ablation Study.}
\subsubsection{Strategy}
This RQ studies how \tool's major design choices contribute to effectiveness.
According to the framework in Figure~\ref{fig:framework}, we focus on two key components.
The first component is Backward Exploit Goal Inference in Layer-wise Execution Plan Construction, which propagates the sink-level requirement backward and constructs layer-wise exploit goals before exploit-state resolution.
To evaluate its contribution, we consider a variant that removes Backward Exploit Goal Inference while retaining Execution Plan Construction. In this variant, \tool still constructs an entry-to-sink execution plan from the path-specific vulnerability-related CPG, but the plan does not contain backward-derived exploit goals. It is then passed to the LLM-Guided Exploit State Resolution module without layer-wise exploit-goal constraints.

The second component is the context selector used in LLM-Guided Exploit State Resolution. 
We keep the rest of VulnSEER unchanged and replace the original RL-based selector with several alternatives. 
The CodeBERT selector is a supervised binary classifier trained on the same selection data. 
The LoRA selector uses the same Qwen3-8B backbone and the same data as the original selector, but is trained only with LoRA fine-tuning. 
The GPT-5.2 and Qwen3.5 selectors directly use the corresponding LLMs to decide which contexts should be included. 
The All-Include variant removes selection and includes all candidate contexts. 
For the ablation study, we report results from one fixed run for all variants.

\begin{table}[t]
\centering
\caption{Ablation results for goal inference and context selection.}
\label{tab:rq3-selector-ablation}
\scriptsize
\setlength{\tabcolsep}{1.2pt}
\renewcommand{\arraystretch}{1.10}
\resizebox{\columnwidth}{!}{
\begin{tabular}{@{}lcccccccc@{}}
\toprule
\multirow{3}{*}{\textbf{Cat.}}
& \multirow{3}{*}{\textbf{GT}}
& \multirow{2}{*}{\textbf{\tool}}
& \multirow{2}{*}{\textbf{w/o BGI}}
& \multicolumn{5}{c}{\textbf{Selectors}} \\
\cmidrule(lr){5-9}
& 
& 
& 
& \textbf{CBERT}
& \textbf{LoRA}
& \textbf{GPT}
& \textbf{Qwen}
& \textbf{All} \\
\cmidrule(lr){3-3}
\cmidrule(lr){4-4}
\cmidrule(lr){5-5}
\cmidrule(lr){6-6}
\cmidrule(lr){7-7}
\cmidrule(lr){8-8}
\cmidrule(lr){9-9}
&
&
\textbf{SR}
& \textbf{SR}
& \textbf{SR}
& \textbf{SR}
& \textbf{SR}
& \textbf{SR}
& \textbf{SR} \\
\midrule

\textbf{Direct}
& \textbf{78}
& \textbf{98.72\%}
& \textbf{98.72\%}
& 96.15\%
& 96.15\%
& \textbf{98.72\%}
& \textbf{98.72\%}
& \textbf{98.72\%} \\

\textbf{Short}
& \textbf{173}
& 84.97\%
& 42.77\%
& 62.43\%
& 65.90\%
& \textbf{87.28\%}
& 80.35\%
& 83.82\% \\

\textbf{Long}
& \textbf{18}
& \textbf{66.67\%}
& 11.11\%
& 22.22\%
& 27.78\%
& 50.00\%
& 50.00\%
& 11.11\% \\

\midrule
\textbf{Total}
& \textbf{269}
& 87.73\%
& 56.88\%
& 69.52\%
& 72.12\%
& \textbf{88.10\%}
& 83.64\%
& 83.27\% \\

\bottomrule
\end{tabular}}
\vspace{1mm}

\parbox{\columnwidth}{\scriptsize \textit{Note.} BGI denotes backward goal inference; CBERT denotes CodeBERT.}
\vspace{-5mm}
\end{table}

\subsubsection{Results}
Table~\ref{tab:rq3-selector-ablation} evaluates both components under the same downstream resolver.
Backward goal inference is critical to \tool's effectiveness. 
Removing it drops the overall success rate from 87.73\% (236/269) to 56.88\% (153/269). 
The degradation is most pronounced on non-trivial chains: short-chain success drops from 84.97\% to 42.77\% (73 fewer cases), and long-chain success drops from 66.67\% to 11.11\% (only 2 of 18 cases). 
This confirms that backward goal inference supplies essential intermediate constraints for guiding exploit construction across caller-callee transitions.

The context selector also substantially affects effectiveness. 
The RL-based selector outperforms CodeBERT and LoRA, which succeed in only 187 and 194 cases respectively, indicating that supervised classification or lightweight fine-tuning alone cannot consistently identify exploit-relevant context. 
This advantage grows with chain complexity: on short chains, the RL-based selector succeeds in 147/173 cases versus 108 and 114 for CodeBERT and LoRA; on long chains, it solves 12/18 cases versus only 4, 5, and 2 for CodeBERT, LoRA, and All-Include. 
All-Include performs worst on long chains despite retaining all candidate context, suggesting that excessive context induces context explosion and distracts the resolver at critical branch points.

Compared with LLM-based selectors, the RL-based selector offers a more favorable trade-off on longer chains. 
The GPT-5.2 selector achieves the highest overall count (237 cases) with slightly better short-chain performance, but solves only 9 long-chain cases, fewer than the RL-based selector's 12. 
The Qwen3.5 selector similarly achieves 225 cases overall but only 9 on long chains. 
These results indicate that \tool's advantage comes not from adding more context or relying on stronger LLM judgment, but from selecting a compact, goal-relevant context set that preserves key state constraints across deeper call chains.

\begin{tcolorbox}
\textbf{Answer to RQ2:} 
Backward goal inference and RL-based context selection are both essential to \tool.
The former makes exploit construction goal-directed, while the latter helps the resolver focus on compact, exploit-relevant contexts.
Together, they reduce reasoning deviation and context explosion, improving vulnerability exposure on longer call chains.
\end{tcolorbox}

\subsection{RQ3: Efficiency and Cost.}
\begin{table*}[t]
\centering
\caption{Efficiency comparison under different context-selection variants and baseline.}
\label{tab:rq4-efficiency-cost}
\scriptsize
\setlength{\tabcolsep}{2.0pt}
\renewcommand{\arraystretch}{1.12}
\resizebox{\textwidth}{!}{
\begin{tabular}{l|ccc|*{15}{c}|cc}
\toprule
\multirow{3}{*}{\textbf{Category}}
& \multicolumn{3}{c|}{\multirow{2}{*}{\textbf{\tool}}}
& \multicolumn{15}{c|}{\textbf{Selectors}}
& \multicolumn{2}{c}{\multirow{2}{*}{\textbf{\textsc{Magneto}}}} \\
\cmidrule(lr){5-19}
&
& 
& 
& \multicolumn{3}{c}{\makecell{\textbf{CodeBERT}}}
& \multicolumn{3}{c}{\makecell{\textbf{LoRA}}}
& \multicolumn{3}{c}{\makecell{\textbf{GPT-5.2}}}
& \multicolumn{3}{c}{\makecell{\textbf{Qwen3.5}}}
& \multicolumn{3}{c|}{\textbf{All-Include}}
& 
&  \\
\cmidrule(lr){2-4}
\cmidrule(lr){5-7}
\cmidrule(lr){8-10}
\cmidrule(lr){11-13}
\cmidrule(lr){14-16}
\cmidrule(lr){17-19}
\cmidrule(lr){20-21}
& \textbf{Ctx.} & \textbf{Time(s)} & \textbf{Tokens}
& \textbf{Ctx.} & \textbf{Time(s)} & \textbf{Tokens}
& \textbf{Ctx.} & \textbf{Time(s)} & \textbf{Tokens}
& \textbf{Ctx.} & \textbf{Time(s)} & \textbf{Tokens}
& \textbf{Ctx.} & \textbf{Time(s)} & \textbf{Tokens}
& \textbf{Ctx.} & \textbf{Time(s)} & \textbf{Tokens}
& \textbf{Time(s)} & \textbf{Tokens} \\
\midrule

\textbf{Direct}
& 1.73 & 114.88 & 7,884
& 0.05 & {95.58} & {6,284}
& 0.32 & 133.36 & 6,752
& 1.79 & 277.98 & 11,936
& 0.96 & 332.47 & 11,528
& 3.06 & \textbf{94.76} & 9,424
& 576.91 & \textbf{3,473} \\

\textbf{Short}
& 6.85 & 197.81 & 28,857
& 0.46 & \textbf{119.71} & {24,178}
& 3.18 & 190.14 & 26,159
& 7.09 & 640.99 & 47,030
& 5.37 & 786.24 & 45,583
& 13.28 & {119.91} & 35,135
& 1254.51 & \textbf{9,842} \\

\textbf{Long}
& 18.59 & 257.95 & 105,943
& 0.61 & {188.61} & {74,392}
& 10.50 & 275.32 & 86,955
& 19.28 & 1793.91 & 156,974
& 17.44 & 2227.38 & 150,083
& 35.89 & \textbf{181.87} & 141,972
& 3624.85 & \textbf{28,759} \\

\midrule
\textbf{Average}
& 6.15 & 177.79 & 27,934
& 0.35 & {117.33} & {22,350}
& 2.84 & 179.37 & 24,600
& 6.37 & 612.88 & 44,211
& 4.90 & 751.10 & 42,701
& 11.83 & \textbf{116.77} & 34,829
& 1436.93 & \textbf{11,019} \\

\bottomrule
\end{tabular}} \vspace{-3mm}
\end{table*}

\subsubsection{Strategy}
% This RQ evaluates the efficiency and cost of VulnSEER. 
We measure end-to-end execution time and token cost for the whole framework. 
Since context selection affects both resolver input size and model-invocation count, we compare \tool with selector variants used in RQ2, including CodeBERT Selector, LoRA Selector, GPT-5.2 Selector, Qwen3.5 Selector, and All-Include. We also compare with \textsc{Magneto} to assess the trade-off among execution time, token cost, and exploit-generation effectiveness.
We report results based on a single run for each evaluated configuration.

\subsubsection{Results}
Table~\ref{tab:rq4-efficiency-cost} reports the average execution time and token cost across selector variants. 
\tool selects 6.15 context evidences on average from 11.83 candidates. 
CodeBERT and LoRA select only 0.35 and 2.84 evidences, respectively, consuming fewer tokens but over-filtering exploit-relevant context, which explains their lower success rates in RQ2, particularly on short and long chains. 
All-Include retains all candidates, avoiding omission but raising token cost to 34,829 and degrading long-chain effectiveness.

Compared with LLM-based selectors, \tool is substantially more efficient despite selecting a similar number of contexts. 
\tool requires 177.79 seconds on average, versus 612.88 and 751.10 seconds for the GPT-5.2 and Qwen3.5 selectors. 
On long chains, the GPT-5.2 and Qwen3.5 selectors take 1793.91 and 2227.38 seconds, respectively, which are 7.0 and 8.6 times slower than \tool. 
This stems from the additional model invocations LLM-based selectors require as the candidate pool grows, which also raises their token cost to 44,211 and 42,701, compared with 27,934 for \tool.

Compared with \textsc{Magneto}, \tool consumes more tokens (27,934 vs.\ 11,019) but is over 7 times faster overall (177.79 vs.\ 1436.93 seconds), with the gap most pronounced on long chains (257.95 vs.\ 3624.85 seconds). 
Given \tool's substantially higher success rate from RQ1, the added token cost translates into more effective and faster exploit generation rather than overhead.

\begin{tcolorbox}
\textbf{Answer to RQ3:}
The RL-based selector gives \tool a favorable efficiency-cost trade-off, keeping token consumption low while maintaining high execution time efficiency through compact, goal-relevant context selection.
\end{tcolorbox}

%区分出长链和短链，画一个其他baseline的折线图，可以从斜率最大的点作为区分长短链的依据，这需要在intro里就体现出来；rq1需要的是定量分析，从数据的角度分长短链分开描述我们的实验结果，表格可以把不同LLM的实验数据带上，比如vulnseer-gpt5.2；rq2是定性分析，分长短链画出维恩图，重点说明长链中我们单独可以解决的case，重点是要突出我们解决这些case是因为解决了我们intro中提出的挑战。motivating example的case要和rq2中的case是一样的。

\section{Related Work}
%\xiaoxue{[Rewritten, please double-check]}
\subsection{Vulnerable Dependency Detection and Reachability.}

Research on third-party library security mainly focuses on vulnerability detection, which identifies vulnerable dependencies or code patterns~\cite{pashchenko2018vulnerable,pashchenko2022vuln4real,imtiaz2021comparative,yamaguchi2014modeling,li2018vuldeepecker,zhou2019devign,fu2022linevul}, and reachability analysis, which determines whether vulnerable functions are callable from client entry points~\cite{ponta2018beyond,ponta2020detection,mir2023effect,sun2023silent}. While these approaches reduce false positives, they do not determine whether a client entry can construct the runtime state required to trigger the vulnerability. \tool addresses this gap by synthesizing and dynamically validating entry-side exploit payloads for recovered call chains.

%Research on third-party library security has progressed along two lines.
%Vulnerability detection approaches identify whether a project depends on vulnerable versions or contains vulnerable code patterns via metadata matching, API signature comparison, patch analysis, and learned code representations~\cite{pashchenko2018vulnerable,pashchenko2022vuln4real,imtiaz2021comparative,zhao2023sca,yamaguchi2014modeling,li2018vuldeepecker,li2021sysevr,zhou2019devign,fu2022linevul}.
%Reachability analysis further determines whether vulnerable functions are transitively callable from client entry points via call graph construction and usage-based pruning~\cite{ponta2018beyond,ponta2020detection,mir2023effect,zhang2024vascanner,sun2023silent}.
%Both lines reduce false positives over package-level alerts, but neither establishes whether a client entry can construct the concrete runtime state required to trigger the vulnerability.
%\tool targets this residual gap by synthesizing and dynamically validating entry-side exploit payloads for statically recovered call chains.

\subsection{Vulnerability Exploit Generation.}
Moving beyond reachability, other works generate executable evidence that clients can trigger vulnerable library behavior. 
Traditional automated exploit generation focused on binary vulnerabilities via patch differencing, constraint solving, and binary analysis~\cite{brumley2008automatic,avgerinos2011aeg,cha2012mayhem,shoshitaishvili2016sok}.
Recent approaches include evolutionary search to reproduce library-side vulnerability states~\cite{kang2022test}, exploit-value migration from public PoCs~\cite{chen2024exploiting}, caller--callee decomposition for fuzzing guidance~\cite{zhou2024magneto}, and agent-based exploit reproduction using vulnerability reports and iterative repair~\cite{iannone2021toward,gao2025vulnerability,zhao2026triggering,nitin2025faultline,simsek2026pocgen,chen2026multiagent,liu2026dual,ni2026taint}. These methods largely treat exploit synthesis as propagating triggering values through the call chain. However, on long chains the required condition is often re-expressed across successive boundaries as receiver fields, container elements, or factory decisions rather than remaining a direct argument. \tool instead models exploit synthesis as \textit{state re-expression}, inferring how sink-level requirements map to caller-controllable state and selecting the off-chain context needed to realize them.

%To move beyond reachability, recent work generates executable evidence that a client can trigger vulnerable library behavior.
%Traditional automated exploit generation focused on binary vulnerabilities via patch differencing, constraint solving, and binary analysis~\cite{brumley2008automatic,avgerinos2011aeg,cha2012unleashing,shoshitaishvili2016sok}.
%For vulnerable dependencies, TRANSFER~\cite{kang2022test} reproduces a state resembling the library-side vulnerability witness via evolutionary search, VESTA~\cite{chen2024vesta} migrates concrete triggering values from public exploits into reachable client tests via transformation rules, and MAGNETO~\cite{zhou2024magneto} decomposes a chain into caller-callee steps and generates reverse-direction seeds to guide fuzzing.
%More recent agent-based systems incorporate vulnerability reports and iterative repair to reproduce PoC exploits within a known environment~\cite{iannone2021toward,gao2025vuleut,zhao2026livefuzz,nitin2025faultline,simsek2025pocgen,chen2026vulnsage,liu2026cve2poc,ni2026llmvd}.
%All these approaches treat exploit synthesis as value propagation: carry a triggering value through the chain.
%This assumption breaks on long chains where the triggering condition is not a direct argument but is reified as receiver fields, container elements, or factory decisions at successive boundaries.
%\tool reframes this as \textit{state re-expression}: at each layer it infers how the sink-level requirement maps onto caller-controllable state and selects the off-chain construction context needed to realize it.

\subsection{Exploit-State Reasoning and Context Analysis.}
Symbolic and concolic execution~\cite{king1976symbolic,godefroid2005dart,sen2005cute,cadar2006exe,cadar2013symbolic,baldoni2018survey}, extensions to heap-manipulating programs~\cite{khurshid2003generalized}, and grammar-structured inputs~\cite{godefroid2008grammar} provide the foundation for constraint-based exploit-state reasoning, with recent LLM-assisted variants alleviating structured input generation~\cite{li2025llmse,wang2024python}.
Program slicing~\cite{horwitz1990interprocedural,tip1995survey}, thin slicing~\cite{sridharan2007thin}, directed fuzzing, and selective symbolic execution~\cite{bohme2017directed,chen2018hawkeye,stephens2016driller,chipounov2011s2e} narrow analysis scope toward a target.
%Yet, these techniques fix a slicing criterion at the sink or assume construction context is recoverable along the chain.
These two lines of work provide the foundation for \tool, but they are typically studied separately in general testing, path exploration, or scope reduction settings.
Dependency exploitation requires a sink-level condition to be re-expressed through caller-side state carriers before it becomes a concrete client-side payload, while relevant context evolves with the current exploit goal as the sink condition moves across arguments, fields, containers, factories, configuration keys, and dynamic-dispatch targets.
\tool addresses this challenge by combining backward exploit-goal inference, which re-expresses the sink-level condition layer by layer, with an RL-based context selector that identifies the off-chain code governing each state transition.

\section{Threats to Validity}

\noindent\textbf{Internal Validity.}
Our benchmark extends the \textsc{Magneto} dataset with additional exploitable call chains, each manually validated by two Java security experts using dynamic oracles, reducing false positives from static reachability. Although oracle fidelity and implementation differences may affect results, we mitigate LLM nondeterminism by fixing model versions, decoding parameters, retry budgets, and timeouts across tools.

%Our benchmark extends the \textsc{Magneto} dataset with additional exploitable call chains, each manually inspected and confirmed by two Java security experts through dynamic oracle execution. A payload is accepted only when it reaches the vulnerable sink and satisfies the vulnerability-specific oracle, reducing false positives from purely static reachability. Oracle implementations may nonetheless be imperfect for vulnerabilities with complex runtime effects or environment-dependent behavior. To mitigate LLM nondeterminism, we fix model versions, decoding parameters, retry budgets, and timeouts across all tools, though differences in prompt formats and implementation assumptions may still influence comparisons.

\noindent\textbf{External Validity.}
Our evaluation focuses on Java projects with vulnerable third-party dependencies. The results may not generalize to other languages, ecosystems, vulnerability classes, or complex scenarios. 
%RB-Comment 2：Small Number of Long-chain Cases
The evaluation also includes only 18 long-chain cases, limiting the generalizability of our long-chain results. Expanding this subset through manual validation requires substantial human effort, so we leave this extension for future work. 
We do not probe the impact of data leakage to LLMs. Although public vulnerability reports and exploit sketches may appear in LLM pretraining data, our ablation shows that LLM-only and alternative selector variants still fail on long chains, suggesting that memorized artifacts alone cannot preserve exploit conditions across project-specific state transformations.

%Our evaluation covers Java projects with vulnerable third-party dependencies across direct, short, and long call chains. The results may not generalize to other languages, ecosystems, or vulnerability classes, nor to scenarios involving reflection-heavy dispatch, asynchronous execution, or external attacker-controlled state. Regarding data leakage, public vulnerability reports and exploit sketches may appear in LLM pretraining data; however, our ablation shows that LLM-only and alternative selector variants still fail on long chains, indicating that memorized artifacts alone are insufficient to preserve exploit conditions across project-specific state transformations.

\section{Conclusion}
This paper presents \tool, an LLM-guided framework for synthesizing vulnerability-triggering exploit payloads in Java projects with vulnerable third-party dependencies. \tool performs backward exploit-goal inference to construct a layer-wise execution plan, resolves exploit states with an RL-based context selector, and validates synthesized payloads through dynamic checking. Evaluated on 269 ground-truth exploitable call chains, \tool consistently outperforms existing baselines, especially on non-direct chains, demonstrating the effectiveness of exploit-goal planning and context-guided exploit-state resolution.

%This paper presents \tool, an LLM-guided exploit-state synthesis framework for synthesizing vulnerability-reaching exploit payloads in Java projects with vulnerable third-party dependencies. \tool first prepares vulnerability-related information by constructing path-specific CPGs and extracting sink-level requirements, then performs layer-wise execution plan construction through backward exploit goal inference. Guided by the resulting execution plan, \tool resolves exploit states with an RL-based exploit-state context selector and validates synthesized payloads through dynamic exploitability checking. On 269 ground-truth exploitable call chains, \tool outperforms all baselines, especially on non-direct chains. These results show that combining vulnerability-related information preparation, backward exploit-goal planning, and context-guided exploit-state resolution can improve practical exploitability assessment for complex vulnerable-dependency paths.

\section{Data Availability}
The replication package for this study, including the 269 evaluation cases, their corresponding ground-truth call chains and vulnerability oracles, prompt templates, and documentation, will be publicly available upon publication at \url{https://github.com/Ylkylkylk/VulnSEER}. 

\section*{Acknowledgments}
This work is supported by the Yongjiang Talent Program (No. 2023A-402-G), the National Natural Science Foundation of China (No. 62302437, No. 62372398, No. 72342025), the Zhejiang Pioneer (Jianbing) Project (2025C01198), Zhejiang Provincial Science Foundation of China (No. LQK26F020002), and the Fundamental Research Funds for the Central Universities (No. 226-2025-00067).
% =========================================================
% Bibliography
% =========================================================
\bibliographystyle{IEEEtran}
\bibliography{references}

@inproceedings{deng2025chainfuzz,
  title={$\{$ChainFuzz$\}$: Exploiting upstream vulnerabilities in $\{$Open-Source$\}$ supply chains},
  author={Deng, Peng and Zhang, Lei and Meng, Yuchuan and Yang, Zhemin and Zhang, Yuan},
  booktitle={34th USENIX Security Symposium (USENIX Security 25)},
  pages={6199--6218},
  year={2025}
}

@inproceedings{sun2023silent,
  author        = {Sun, J. and Xing, Z. and Lu, Q. and Xu, X. and Zhu, L. and Hoang, T. and Zhao, D.},
  title         = {{Silent Vulnerable Dependency Alert Prediction with Vulnerability Key Aspect Explanation}},
  booktitle     = {{2023 IEEE/ACM 45th International Conference on Software Engineering (ICSE)}},
  year          = {2023},
  pages         = {970--982},
  publisher     = {IEEE},
  doi           = {10.1109/ICSE48619.2023.00089},
  url           = {https://doi.org/10.1109/ICSE48619.2023.00089}
}

@article{ponta2020detection,
  author        = {Ponta, S. E. and Plate, H. and Sabetta, A.},
  title         = {{Detection, Assessment and Mitigation of Vulnerabilities in Open Source Dependencies}},
  journal       = {{Empirical Software Engineering}},
  year          = {2020},
  volume        = {25},
  number        = {5},
  pages         = {3175--3215},
  doi           = {10.1007/s10664-020-09830-x},
  url           = {https://doi.org/10.1007/s10664-020-09830-x}
}

@inproceedings{mir2023effect,
  author        = {Mir, A. M. and Keshani, M. and Proksch, S.},
  title         = {{On the Effect of Transitivity and Granularity on Vulnerability Propagation in the Maven Ecosystem}},
  booktitle     = {{2023 IEEE International Conference on Software Analysis, Evolution and Reengineering (SANER)}},
  year          = {2023},
  pages         = {201--211},
  publisher     = {IEEE},
  doi           = {10.1109/SANER56733.2023.00028},
  url           = {https://doi.org/10.1109/SANER56733.2023.00028}
}

@article{pashchenko2022vuln4real,
  author        = {Pashchenko, I. and Plate, H. and Ponta, S. E. and Sabetta, A. and Massacci, F.},
  title         = {{Vuln4Real: A Methodology for Counting Actually Vulnerable Dependencies}},
  journal       = {{IEEE Transactions on Software Engineering}},
  year          = {2022},
  volume        = {48},
  number        = {5},
  pages         = {1592--1609},
  doi           = {10.1109/TSE.2020.3025443},
  url           = {https://doi.org/10.1109/TSE.2020.3025443}
}

@inproceedings{zhao2023software,
  author        = {Zhao, L. and Chen, S. and Xu, Z. and Liu, C. and Zhang, L. and Wu, J. and Sun, J. and Liu, Y.},
  title         = {{Software Composition Analysis for Vulnerability Detection: An Empirical Study on Java Projects}},
  booktitle     = {{Proceedings of the 31st ACM Joint European Software Engineering Conference and Symposium on the Foundations of Software Engineering}},
  series        = {{ESEC/FSE '23}},
  year          = {2023},
  pages         = {960--972},
  publisher     = {ACM},
  doi           = {10.1145/3611643.3616299},
  url           = {https://doi.org/10.1145/3611643.3616299}
}

@inproceedings{imtiaz2021comparative,
  author        = {Imtiaz, N. and Thorne, S. and Williams, L.},
  title         = {{A Comparative Study of Vulnerability Reporting by Software Composition Analysis Tools}},
  booktitle     = {{Proceedings of the 15th ACM/IEEE International Symposium on Empirical Software Engineering and Measurement}},
  series        = {{ESEM '21}},
  year          = {2021},
  publisher     = {ACM},
  pages         = {1--11},
  doi           = {10.1145/3475716.3475769},
  url           = {https://doi.org/10.1145/3475716.3475769}
}

@article{li2022sysevr,
  author        = {Li, Z. and Zou, D. and Xu, S. and Jin, H. and Zhu, Y. and Chen, Z.},
  title         = {{SySeVR: A Framework for Using Deep Learning to Detect Software Vulnerabilities}},
  journal       = {{IEEE Transactions on Dependable and Secure Computing}},
  year          = {2022},
  volume        = {19},
  number        = {4},
  pages         = {2244--2258},
  doi           = {10.1109/TDSC.2021.3051525},
  url           = {https://doi.org/10.1109/TDSC.2021.3051525}
}

@inproceedings{fu2022linevul,
  author        = {Fu, M. and Tantithamthavorn, C.},
  title         = {{LineVul: A Transformer-based Line-Level Vulnerability Prediction}},
  booktitle     = {{Proceedings of the 19th International Conference on Mining Software Repositories}},
  series        = {{MSR '22}},
  year          = {2022},
  pages         = {608--620},
  publisher     = {ACM},
  doi           = {10.1145/3524842.3528452},
  url           = {https://doi.org/10.1145/3524842.3528452}
}

@inproceedings{ponta2018beyond,
  author        = {Ponta, S. E. and Plate, H. and Sabetta, A.},
  title         = {{Beyond Metadata: Code-Centric and Usage-Based Analysis of Known Vulnerabilities in Open-Source Software}},
  booktitle     = {{2018 IEEE International Conference on Software Maintenance and Evolution (ICSME)}},
  year          = {2018},
  pages         = {449--460},
  publisher     = {IEEE},
  doi           = {10.1109/ICSME.2018.00054},
  url           = {https://doi.org/10.1109/ICSME.2018.00054}
}

@article{zhang2024does,
  author        = {Zhang, F. and Fan, L. and Chen, S. and Cai, M. and Xu, S. and Zhao, L.},
  title         = {{Does the Vulnerability Threaten Our Projects? Automated Vulnerable API Detection for Third-Party Libraries}},
  journal       = {{IEEE Transactions on Software Engineering}},
  year          = {2024},
  volume        = {50},
  number        = {11},
  pages         = {2906--2920},
  doi           = {10.1109/TSE.2024.3454960},
  url           = {https://doi.org/10.1109/TSE.2024.3454960}
}

@article{jia2025impact,
  author        = {Jia, Z. and Yang, C. and Feng, P. and Zhao, X. and Li, X. and Ma, J.},
  title         = {{Impact Assessment of Third-Party Library Vulnerabilities Through Vulnerability Reachability Analysis}},
  journal       = {{Computers \& Security}},
  year          = {2025},
  volume        = {157},
  pages         = {104546},
  doi           = {10.1016/j.cose.2025.104546},
  url           = {https://doi.org/10.1016/j.cose.2025.104546}
}

@misc{wang2025savant,
  author        = {Wang, L. and Fu, Q. and Song, W. and Deng, G. and Liu, Y. and Williams, D. and Zhang, Y.},
  title         = {{SAVANT: Vulnerability Detection in Application Dependencies Through Semantic-Guided Reachability Analysis}},
  year          = {2025},
  eprint        = {2506.17798},
  archivePrefix = {arXiv},
  doi           = {10.48550/arXiv.2506.17798},
  url           = {https://arxiv.org/abs/2506.17798}
}

@article{ma2024vulnet,
  author        = {Ma, Z. and Mondal, S. and Chen, T.-H. and Zhang, H. and Hassan, A. E.},
  title         = {{VulNet: Towards Improving Vulnerability Management in the Maven Ecosystem}},
  journal       = {{Empirical Software Engineering}},
  year          = {2024},
  volume        = {29},
  number        = {4},
  articleno     = {83},
  doi           = {10.1007/s10664-024-10448-6},
  url           = {https://doi.org/10.1007/s10664-024-10448-6}
}

@inproceedings{kang2022test,
  author        = {Kang, H. J. and Nguyen, T. G. and Le, B. and Pasareanu, C. S. and Lo, D.},
  title         = {{Test Mimicry to Assess the Exploitability of Library Vulnerabilities}},
  booktitle     = {{Proceedings of the 31st ACM SIGSOFT International Symposium on Software Testing and Analysis}},
  series        = {{ISSTA '22}},
  year          = {2022},
  pages         = {276--288},
  publisher     = {ACM},
  doi           = {10.1145/3533767.3534398},
  url           = {https://doi.org/10.1145/3533767.3534398}
}

@inproceedings{chen2024exploiting,
  author        = {Chen, Z. and Hu, X. and Xia, X. and Gao, Y. and Xu, T. and Lo, D. and Yang, X.},
  title         = {{Exploiting Library Vulnerability via Migration Based Automating Test Generation}},
  booktitle     = {{Proceedings of the IEEE/ACM 46th International Conference on Software Engineering}},
  series        = {{ICSE '24}},
  year          = {2024},
  pages         = {1--12},
  publisher     = {ACM},
  doi           = {10.1145/3597503.3639583},
  url           = {https://doi.org/10.1145/3597503.3639583}
}

@inproceedings{zhou2024magneto,
  author        = {Zhou, Z. and Yang, Y. and Wu, S. and Huang, Y. and Chen, B. and Peng, X.},
  title         = {{Magneto: A Step-Wise Approach to Exploit Vulnerabilities in Dependent Libraries via LLM-Empowered Directed Fuzzing}},
  booktitle     = {{Proceedings of the 39th IEEE/ACM International Conference on Automated Software Engineering}},
  series        = {{ASE '24}},
  year          = {2024},
  pages         = {1633--1644},
  publisher     = {ACM},
  doi           = {10.1145/3691620.3695531},
  url           = {https://doi.org/10.1145/3691620.3695531}
}

@article{baldoni2018survey,
  author        = {Baldoni, R. and Coppa, E. and D'Elia, D. C. and Demetrescu, C. and Finocchi, I.},
  title         = {{A Survey of Symbolic Execution Techniques}},
  journal       = {{ACM Computing Surveys}},
  year          = {2018},
  volume        = {51},
  number        = {3},
  pages         = {1--39},
  doi           = {10.1145/3182657},
  url           = {https://doi.org/10.1145/3182657}
}

@inproceedings{tu2026cottontail,
  author        = {Tu, H. and Lee, S. and Li, Y. and Chen, P. and Jiang, L. and Bohme, M.},
  title         = {{Cottontail: Large Language Model-Driven Concolic Execution for Highly Structured Test Input Generation}},
  booktitle     = {{2026 IEEE Symposium on Security and Privacy (SP)}},
  year          = {2026},
  pages         = {3509--3527},
  publisher     = {IEEE},
  doi           = {10.1109/SP63933.2026.00110},
  url           = {https://doi.org/10.1109/SP63933.2026.00110}
}

@inproceedings{sridharan2007thin,
  author        = {Sridharan, M. and Fink, S. J. and Bodik, R.},
  title         = {{Thin Slicing}},
  booktitle     = {{Proceedings of the ACM SIGPLAN Conference on Programming Language Design and Implementation}},
  series        = {{PLDI '07}},
  year          = {2007},
  pages         = {112--122},
  publisher     = {ACM},
  doi           = {10.1145/1273442.1250748},
  url           = {https://doi.org/10.1145/1273442.1250748}
}

@misc{chen2026multiagent,
  author        = {Chen, S. and Luo, T. and Wu, S. and Liu, X. and Zhou, Y. and Li, Q. and Xu, W.},
  title         = {{A Multi-Agent Framework for Automated Exploit Generation with Constraint-Guided Comprehension and Reflection}},
  year          = {2026},
  eprint        = {2604.05130},
  archivePrefix = {arXiv},
  doi           = {10.48550/arXiv.2604.05130},
  url           = {https://arxiv.org/abs/2604.05130}
}

@article{marques2025automated,
  author        = {Marques, F. and Ferreira, M. and Nascimento, A. and Coimbra, M. E. and Santos, N. and Jia, L. and Santos, J. F.},
  title         = {{Automated Exploit Generation for Node.js Packages}},
  journal       = {{Proceedings of the ACM on Programming Languages}},
  year          = {2025},
  volume        = {9},
  number        = {PLDI},
  pages         = {1341--1366},
  doi           = {10.1145/3729304},
  url           = {https://doi.org/10.1145/3729304}
}

@misc{joern2024workbench,
  author        = {{Joern Project}},
  title         = {{Joern: The Bug Hunter's Workbench}},
  year          = {2024},
  month         = {jan},
  howpublished  = {{GitHub repository}},
  url           = {https://github.com/joernio/joern},
  note          = {Accessed: July 7, 2026.}
}

@article{hoare1969axiomatic,
  author        = {Hoare, C. A. R.},
  title         = {{An Axiomatic Basis for Computer Programming}},
  journal       = {{Communications of the ACM}},
  year          = {1969},
  volume        = {12},
  number        = {10},
  pages         = {576--580},
  doi           = {10.1145/363235.363259},
  url           = {https://doi.org/10.1145/363235.363259}
}

@misc{asm2024,
  author        = {{OW2 Consortium}},
  title         = {{ASM}},
  year          = {2024},
  howpublished  = {{Project website}},
  url           = {https://asm.ow2.io/},
  note          = {Accessed: July 7, 2026.}
}

@inproceedings{pashchenko2018vulnerable,
  author        = {Pashchenko, I. and Plate, H. and Ponta, S. E. and Sabetta, A. and Massacci, F.},
  title         = {{Vulnerable Open Source Dependencies: Counting Those That Matter}},
  booktitle     = {{Proceedings of the 12th ACM/IEEE International Symposium on Empirical Software Engineering and Measurement}},
  series        = {{ESEM '18}},
  year          = {2018},
  pages         = {1--10},
  publisher     = {ACM},
  doi           = {10.1145/3239235.3268920},
  url           = {https://doi.org/10.1145/3239235.3268920}
}

@inproceedings{yamaguchi2014modeling,
  author        = {Yamaguchi, F. and Golde, N. and Arp, D. and Rieck, K.},
  title         = {{Modeling and Discovering Vulnerabilities with Code Property Graphs}},
  booktitle     = {{2014 IEEE Symposium on Security and Privacy}},
  year          = {2014},
  pages         = {590--604},
  publisher     = {IEEE},
  doi           = {10.1109/SP.2014.44},
  url           = {https://doi.org/10.1109/SP.2014.44}
}

@inproceedings{li2018vuldeepecker,
  author        = {Li, Z. and Zou, D. and Xu, S. and Ou, X. and Jin, H. and Wang, S. and Deng, Z. and Zhong, Y.},
  title         = {{VulDeePecker: A Deep Learning-Based System for Vulnerability Detection}},
  booktitle     = {{Proceedings of the Network and Distributed System Security Symposium}},
  series        = {{NDSS '18}},
  year          = {2018},
  doi           = {10.14722/ndss.2018.23158},
  url           = {https://doi.org/10.14722/ndss.2018.23158}
}

@inproceedings{zhou2019devign,
  author        = {Zhou, Y. and Liu, S. and Siow, J. and Du, X. and Liu, Y.},
  title         = {{Devign: Effective Vulnerability Identification by Learning Comprehensive Program Semantics via Graph Neural Networks}},
  booktitle     = {{Advances in Neural Information Processing Systems}},
  volume        = {32},
  year          = {2019},
  url           = {https://proceedings.neurips.cc/paper/2019/hash/49265d2447bc3bbfe9e76306ce40a31f-Abstract.html}
}

@inproceedings{brumley2008automatic,
  author        = {Brumley, D. and Poosankam, P. and Song, D. and Zheng, J.},
  title         = {{Automatic Patch-Based Exploit Generation is Possible: Techniques and Implications}},
  booktitle     = {{Proceedings of the IEEE Symposium on Security and Privacy}},
  year          = {2008},
  pages         = {143--157},
  publisher     = {IEEE},
  doi           = {10.1109/SP.2008.17},
  url           = {https://doi.org/10.1109/SP.2008.17}
}

@inproceedings{avgerinos2011aeg,
  author        = {Avgerinos, T. and Cha, S. K. and Hao, B. L. T. and Brumley, D.},
  title         = {{AEG: Automatic Exploit Generation}},
  booktitle     = {{Proceedings of the Network and Distributed System Security Symposium}},
  series        = {{NDSS '11}},
  year          = {2011},
  url           = {https://www.ndss-symposium.org/ndss2011/aeg-automatic-exploit-generation/}
}

@inproceedings{cha2012mayhem,
  author        = {Cha, S. K. and Avgerinos, T. and Rebert, A. and Brumley, D.},
  title         = {{Unleashing Mayhem on Binary Code}},
  booktitle     = {{Proceedings of the IEEE Symposium on Security and Privacy}},
  year          = {2012},
  pages         = {380--394},
  publisher     = {IEEE},
  doi           = {10.1109/SP.2012.31},
  url           = {https://doi.org/10.1109/SP.2012.31}
}

@inproceedings{shoshitaishvili2016sok,
  author        = {Shoshitaishvili, Y. and Wang, R. and Salls, C. and Stephens, N. and Polino, M. and Dutcher, A. and Grosen, J. and Feng, S. and Hauser, C. and Kruegel, C. and Vigna, G.},
  title         = {{SoK: (State of) The Art of War: Offensive Techniques in Binary Analysis}},
  booktitle     = {{Proceedings of the IEEE Symposium on Security and Privacy}},
  year          = {2016},
  pages         = {138--157},
  publisher     = {IEEE},
  doi           = {10.1109/SP.2016.17},
  url           = {https://doi.org/10.1109/SP.2016.17}
}

@inproceedings{iannone2021toward,
  author        = {Iannone, E. and Di Nucci, D. and Sabetta, A. and De Lucia, A.},
  title         = {{Toward Automated Exploit Generation for Known Vulnerabilities in Open-Source Libraries}},
  booktitle     = {{Proceedings of the 29th IEEE/ACM International Conference on Program Comprehension}},
  series        = {{ICPC '21}},
  year          = {2021},
  pages         = {396--400},
  publisher     = {IEEE},
  doi           = {10.1109/ICPC52881.2021.00046},
  url           = {https://doi.org/10.1109/ICPC52881.2021.00046}
}

@inproceedings{gao2025vulnerability,
  author        = {Gao, Y. and Hu, X. and Chen, Z. and Xu, T. and Yang, X.},
  title         = {{Vulnerability-Triggering Test Case Generation from Third-Party Libraries}},
  booktitle     = {{2025 IEEE/ACM Second International Conference on AI Foundation Models and Software Engineering (Forge)}},
  year          = {2025},
  pages         = {125--135},
  publisher     = {IEEE},
  doi           = {10.1109/FORGE66646.2025.00021},
  url           = {https://doi.org/10.1109/FORGE66646.2025.00021}
}

@misc{zhao2026triggering,
  author        = {Zhao, Y. and Wu, M. and Hu, X. and Wang, S. and Luo, M. and Xia, X.},
  title         = {{Triggering and Detecting Exploitable Library Vulnerability from the Client by Directed Greybox Fuzzing}},
  year          = {2026},
  eprint        = {2604.04102},
  archivePrefix = {arXiv},
  doi           = {10.48550/arXiv.2604.04102},
  url           = {https://arxiv.org/abs/2604.04102}
}

@misc{nitin2025faultline,
  author        = {Nitin, V. and Ray, B. and Moghaddam, R. Z.},
  title         = {{FaultLine: Automated Proof-of-Vulnerability Generation Using LLM Agents}},
  year          = {2025},
  eprint        = {2507.15241},
  archivePrefix = {arXiv},
  doi           = {10.48550/arXiv.2507.15241},
  url           = {https://arxiv.org/abs/2507.15241}
}

@article{simsek2026pocgen,
  author        = {Simsek, D. and Eghbali, A. and Pradel, M.},
  title         = {{PoCGen: Generating Proof-of-Concept Exploits for Vulnerabilities in Npm Packages}},
  journal       = {{Proceedings of the ACM on Software Engineering}},
  year          = {2026},
  volume        = {3},
  number        = {FSE},
  pages         = {3887--3908},
  doi           = {10.1145/3808178},
  url           = {https://doi.org/10.1145/3808178},
  eprint        = {2506.04962},
  archivePrefix = {arXiv}
}

@misc{liu2026dual,
  author        = {Liu, B. and Zhao, Y. and Chen, Z. and Xu, G. and Wang, H.},
  title         = {{A Dual-Loop Agent Framework for Automated Vulnerability Reproduction}},
  year          = {2026},
  eprint        = {2602.05721},
  archivePrefix = {arXiv},
  doi           = {10.48550/arXiv.2602.05721},
  url           = {https://arxiv.org/abs/2602.05721}
}

@misc{ni2026taint,
  author        = {Ni, R. and Christodorescu, M. and Jia, L.},
  title         = {{Taint-Style Vulnerability Detection and Confirmation for Node.js Packages Using LLM Agent Reasoning}},
  year          = {2026},
  eprint        = {2604.20179},
  archivePrefix = {arXiv},
  doi           = {10.48550/arXiv.2604.20179},
  url           = {https://arxiv.org/abs/2604.20179}
}

@article{king1976symbolic,
  author        = {King, J. C.},
  title         = {{Symbolic Execution and Program Testing}},
  journal       = {{Communications of the ACM}},
  year          = {1976},
  volume        = {19},
  number        = {7},
  pages         = {385--394},
  doi           = {10.1145/360248.360252},
  url           = {https://doi.org/10.1145/360248.360252}
}

@inproceedings{godefroid2005dart,
  author        = {Godefroid, P. and Klarlund, N. and Sen, K.},
  title         = {{DART: Directed Automated Random Testing}},
  booktitle     = {{Proceedings of the ACM SIGPLAN 2005 Conference on Programming Language Design and Implementation}},
  series        = {{PLDI '05}},
  year          = {2005},
  pages         = {213--223},
  publisher     = {ACM},
  doi           = {10.1145/1065010.1065036},
  url           = {https://doi.org/10.1145/1065010.1065036}
}

@inproceedings{sen2005cute,
  author        = {Sen, K. and Marinov, D. and Agha, G.},
  title         = {{CUTE: A Concolic Unit Testing Engine for C}},
  booktitle     = {{Proceedings of the 10th European Software Engineering Conference Held Jointly with the 13th ACM SIGSOFT International Symposium on Foundations of Software Engineering}},
  series        = {{ESEC/FSE '05}},
  year          = {2005},
  pages         = {263--272},
  publisher     = {ACM},
  doi           = {10.1145/1081706.1081750},
  url           = {https://doi.org/10.1145/1081706.1081750}
}

@inproceedings{cadar2006exe,
  author        = {Cadar, C. and Ganesh, V. and Pawlowski, P. M. and Dill, D. L. and Engler, D. R.},
  title         = {{EXE: Automatically Generating Inputs of Death}},
  booktitle     = {{Proceedings of the 13th ACM Conference on Computer and Communications Security}},
  series        = {{CCS '06}},
  year          = {2006},
  pages         = {322--335},
  publisher     = {ACM},
  doi           = {10.1145/1180405.1180445},
  url           = {https://doi.org/10.1145/1180405.1180445}
}

@article{cadar2013symbolic,
  author        = {Cadar, C. and Sen, K.},
  title         = {{Symbolic Execution for Software Testing: Three Decades Later}},
  journal       = {{Communications of the ACM}},
  year          = {2013},
  volume        = {56},
  number        = {2},
  pages         = {82--90},
  doi           = {10.1145/2408776.2408795},
  url           = {https://doi.org/10.1145/2408776.2408795}
}

@inproceedings{khurshid2003generalized,
  author        = {Khurshid, S. and Pasareanu, C. S. and Visser, W.},
  title         = {{Generalized Symbolic Execution for Model Checking and Testing}},
  booktitle     = {{Proceedings of the 9th International Conference on Tools and Algorithms for the Construction and Analysis of Systems}},
  series        = {{TACAS '03}},
  year          = {2003},
  pages         = {553--568},
  publisher     = {Springer},
  doi           = {10.1007/3-540-36577-X_40},
  url           = {https://doi.org/10.1007/3-540-36577-X_40}
}

@inproceedings{godefroid2008grammar,
  author        = {Godefroid, P. and Kiezun, A. and Levin, M. Y.},
  title         = {{Grammar-Based Whitebox Fuzzing}},
  booktitle     = {{Proceedings of the ACM SIGPLAN Conference on Programming Language Design and Implementation}},
  series        = {{PLDI '08}},
  year          = {2008},
  pages         = {206--215},
  publisher     = {ACM},
  doi           = {10.1145/1379022.1375607},
  url           = {https://doi.org/10.1145/1379022.1375607}
}

@article{li2025llmse,
  author        = {Li, Y. and Meng, R. and Duck, G. J.},
  title         = {{Large Language Model Powered Symbolic Execution}},
  journal       = {{Proceedings of the ACM on Programming Languages}},
  year          = {2025},
  volume        = {9},
  number        = {OOPSLA2},
  pages         = {3148--3176},
  doi           = {10.1145/3763163},
  url           = {https://doi.org/10.1145/3763163}
}

@misc{wang2024python,
  author        = {Wang, W. and Liu, K. and Chen, A. R. and Li, G. and Jin, Z. and Huang, G. and Ma, L.},
  title         = {{Python Symbolic Execution with LLM-Powered Code Generation}},
  year          = {2024},
  eprint        = {2409.09271},
  archivePrefix = {arXiv},
  doi           = {10.48550/arXiv.2409.09271},
  url           = {https://arxiv.org/abs/2409.09271}
}

@article{horwitz1990interprocedural,
  author        = {Horwitz, S. and Reps, T. and Binkley, D.},
  title         = {{Interprocedural Slicing Using Dependence Graphs}},
  journal       = {{ACM Transactions on Programming Languages and Systems}},
  year          = {1990},
  volume        = {12},
  number        = {1},
  pages         = {26--60},
  doi           = {10.1145/77606.77608},
  url           = {https://doi.org/10.1145/77606.77608}
}

@article{tip1995survey,
  author        = {Tip, F.},
  title         = {{A Survey of Program Slicing Techniques}},
  journal       = {{Journal of Programming Languages}},
  year          = {1995},
  volume        = {3},
  number        = {3},
  pages         = {121--189},
  url           = {https://dblp.org/rec/journals/jpl/Tip95}
}

@inproceedings{bohme2017directed,
  author        = {Bohme, M. and Pham, V.-T. and Nguyen, M.-D. and Roychoudhury, A.},
  title         = {{Directed Greybox Fuzzing}},
  booktitle     = {{Proceedings of the 2017 ACM SIGSAC Conference on Computer and Communications Security}},
  series        = {{CCS '17}},
  year          = {2017},
  pages         = {2329--2344},
  publisher     = {ACM},
  doi           = {10.1145/3133956.3134020},
  url           = {https://doi.org/10.1145/3133956.3134020}
}

@inproceedings{chen2018hawkeye,
  author        = {Chen, H. and Xue, Y. and Li, Y. and Chen, B. and Xie, X. and Wu, X. and Liu, Y.},
  title         = {{Hawkeye: Towards a Desired Directed Grey-Box Fuzzer}},
  booktitle     = {{Proceedings of the ACM SIGSAC Conference on Computer and Communications Security}},
  series        = {{CCS '18}},
  year          = {2018},
  pages         = {2095--2108},
  publisher     = {ACM},
  doi           = {10.1145/3243734.3243849},
  url           = {https://doi.org/10.1145/3243734.3243849}
}

@inproceedings{stephens2016driller,
  author        = {Stephens, N. and Grosen, J. and Salls, C. and Dutcher, A. and Wang, R. and Corbetta, J. and Shoshitaishvili, Y. and Kruegel, C. and Vigna, G.},
  title         = {{Driller: Augmenting Fuzzing Through Selective Symbolic Execution}},
  booktitle     = {{Proceedings of the Network and Distributed System Security Symposium}},
  series        = {{NDSS '16}},
  year          = {2016},
  doi           = {10.14722/ndss.2016.23368},
  url           = {https://doi.org/10.14722/ndss.2016.23368}
}

@inproceedings{chipounov2011s2e,
  author        = {Chipounov, V. and Kuznetsov, V. and Candea, G.},
  title         = {{S2E: A Platform for In-Vivo Multi-Path Analysis of Software Systems}},
  booktitle     = {{Proceedings of the 16th International Conference on Architectural Support for Programming Languages and Operating Systems}},
  series        = {{ASPLOS '11}},
  year          = {2011},
  pages         = {265--278},
  publisher     = {ACM},
  doi           = {10.1145/1950365.1950396},
  url           = {https://doi.org/10.1145/1950365.1950396}
}
\balance
\end{document}